\documentclass[fleqn,usenatbib]{mnras}

\usepackage[utf8]{inputenc}

\usepackage{acronym}
\usepackage{ae,aecompl}
\usepackage{booktabs}
\usepackage{dcolumn}
\usepackage{graphicx}
\usepackage{amsmath,amsfonts,amssymb}
\usepackage{mathrsfs}
\usepackage[normalem]{ulem}
\usepackage{epstopdf}

\usepackage[T1]{fontenc}

\let\oldhref\href
\renewcommand{\href}[2]{\oldhref{#1}{\hbox{#2}}}

\newcommand\barparena[1]{\overset{%
   \scriptscriptstyle(-)}{#1}}

\title[Estimation of PNS structure from CCSN neutrinos]{Efficient estimation method for time evolution of proto-neutron star mass and radius from supernova neutrino signal}

\author[H. Nagakura et al.]{
Hiroki Nagakura$^{1}$\thanks{E-mail: hiroki.nagakura@nao.ac.jp},
David Vartanyan$^{2}$,
\\
$^{1}$Division of Science, National Astronomical Observatory of Japan, 2-21-1 Osawa, Mitaka, Tokyo 181-8588, Japan\\
$^{2}$Astronomy Department and Theoretical Astrophysics Center, University of California, Berkeley, CA 94720, USA\\
}

\date{Accepted XXX. Received YYY; in original form ZZZ}

\pubyear{2021}

\begin{document}
\label{firstpage}
\pagerange{\pageref{firstpage}--\pageref{lastpage}}
\maketitle

\begin{abstract}
In this paper we present a novel method to estimate the time evolution of proto-neutron star (PNS) structure from the neutrino signal in core-collapse supernovae (CCSN). Employing recent results of multi-dimensional CCSN simulations, we delve into a relation between total emitted neutrino energy (TONE) and PNS mass/radius, and we find that they are strongly correlated with each other. We fit the relation by simple polynomial functions connecting TONE to PNS mass and radius as a function of time. By combining another fitting function representing the correlation between TONE and cumulative number of event at each neutrino observatory, PNS mass and radius can be retrieved from purely observed neutrino data. We demonstrate retrievals of PNS mass and radius from mock data of neutrino signal, and we assess the capability of our proposed method. While underlining the limitations of the method, we also discuss the importance of the joint analysis with gravitational wave signal. This would reduce uncertainties of parameter estimations in our method, and may narrow down the possible neutrino oscillation model. The proposed method is a very easy and inexpensive computation, which will be useful in real data analysis of CCSN neutrino signal. 
\end{abstract}



\section{Introduction}\label{sec:intro} 

The direct detection of neutrinos emitted from core-collapse supernova (CCSN) SN 1987A opened a new window to study their inner dynamics \citep{1987PhRvL..58.1490H,1987PhRvL..58.1494B}. Neutrino detection from SN 1987A suggests that an energy of $\sim 10^{53} {\rm erg}$ was released via neutrinos within around ten seconds, and the average energy of individual neutrinos was tens of MeV. This indicates that a hot and compact remnant, proto-neutron star (PNS), was born as a result of massive stellar collapse, and then either cooled down or subsequently collapsed to form a black hole within the same time scale. On the other hand, the detailed analyses regarding the CCSN explosion mechanism and characteristics of PNS needed to be postponed until the next nearby CCSN due to low statistics of SN 1987A neutrino data. Currently operating and future planned facilities have capabilities to provide a high-statistics neutrino signal from nearby CCSN; indeed, the scale of neutrino detectors has increased by orders of magnitude since 1987 \citep[see, e.g.,][]{2012ARNPS..62...81S,2016NCimR..39....1M,2018JPhG...45d3002H}. The high-statistics data will reveal the time dependent features in neutrino signal, which will offer sensitive diagnostics for the details of CCSN dynamics.

Inferring CCSN dynamics and the PNS structure from neutrino signal can be done by using theoretical models as templates. In general, this requires deconvolution of multiple physical effects of neutrino emission from the observed data. Crudely speaking, the neutrino emission from CCSN can be divided into two different components: accretion and diffusion. The former is particularly important up to the time of shock revival, in which the progenitor dependent features and the explosion sign would be imprinted. The determination of the accretion component is, however, rather difficult, since the conversion efficiency of mass accretion energy to neutrino emission depends on the structure of post-shock flow, i.e., it is time- and space-dependent, and it would be influenced by multi-dimensional (multi-D) fluid instabilities.

The diffusion component, on the other hand, dominates the neutrino emission in the late post-bounce phase. Much effort has been expended in developing theoretical models of PNS cooling with particular attention to microphysical aspects, including the nuclear equation-of-state (EOS) and weak interactions \citep{1986ApJ...307..178B,1999ApJ...513..780P,2001PhR...354....1Y,2010PhRvL.104y1101H,2012PhRvL.108f1103R,2013ApJS..205....2N,2019ApJ...878...25N,2019ApJ...887..110S,2019MNRAS.484.5162W,2020PhRvC.101b5804F,2021PTEP.2021a3E01S}. The PNS structure may be constrained from the neutrino signal through PNS cooling theory \citep{2017JCAP...11..036G,2018JCAP...12..006G,2021PTEP.2021b3E01M,2021PhRvD.103b3016L}. It should be mentioned, however, that most of the current PNS cooling models adopt either spherically-symmetric or toy models, i.e., the multi-D effects are abandoned \citep[but see, e.g.,][for a phenomenological treatment of PNS convection in spherically symmetric models]{2012PhRvL.108f1103R}. Very recently, it has been revealed that PNS convection changes the diffusion component, boosting the luminosity of heavy leptonic neutrinos ($\nu_x$) \citep{2006ApJ...645..534D,2020MNRAS.492.5764N}, and decreasing the average energy of electron-type neutrinos ($\nu_e$) and their anti-particles ($\bar{\nu}_e$) \citep{2021MNRAS.500..696N}; hence, the effect needs to be incorporated in theoretical models of the neutrino signal. Another caveat in current theoretical models is that the fall-back accretion component is completely neglected, which also potentially alters the neutrino emission. Indeed, we have witnessed fall-back accretion in the late post-bounce phase ($\gtrsim 1$ s) in many multi-D CCSN models \citep[see, e.g.,][]{2006ApJ...640..891Y,2009ApJ...699..409F,2010ApJ...725L.106W,2018ApJ...852L..19C,2019MNRAS.484.3307M,2020MNRAS.495.3751C,2021MNRAS.506.1462N}. These findings suggest that deciphering the neutrino signal requires multi-D CCSN models with taking into account potentially long-lasting accretion components. Otherwise, the data analysis may miss key signatures imprinted in neutrino signal.

In practice, however, it is not always practical to employ directly multi-D CCSN simulations to serve as a template for observations. This is simply because a significant amount of dedicated simulations are required to reproduce the observed neutrino signal. Another obstacle in this approach is that multi-D numerical CCSN models usually suffer from the so-called realization problem, in which small changes in the initial condition lead to qualitatively different outcomes. This issue is associated with the stochastic nature of the turbulence that commonly manifests in the post-shock flows in the CCSN core. At the moment, these limit the use of multi-D CCSN models in data analysis for real (and future) observations of CCSN neutrinos.

In this paper, we address these issues and provide a useful method to incorporate multi-D effects in the neutrino signal analysis. By employing our recent multi-D CCSN models, we inspect the relation between the neutrino signal and PNS structure. We then develop a method by which to estimate the time evolution of the PNS mass and radius from observed neutrino data. To assess the capability of our proposed method, we demonstrate retrievals of PNS mass and radius from ``mock'' neutrino signal observed at Super-Kamiokande \citep[SK,][]{2016APh....81...39A}, Hyper-Kamiokande \citep[HK,][]{2018arXiv180504163H}, the Deep Underground Neutrino Experiment \citep[DUNE,][]{2016arXiv160105471A,2016arXiv160807853A,2020arXiv200806647A}, the Jiangmen Underground Neutrino Observatory \citep[JUNO,][]{2016JPhG...43c0401A}, and IceCube \citep[][]{2011A&A...535A.109A} as representative cases. It should also be noted that, if the gravitational wave (GW) signal is also available in data analysis, it will be used to complement the estimation. In this paper, we discuss how we can take advantage of simultaneous detections of the GW- and neutrino signal to improve estimates of the PNS structure.

\section{Basic idea}\label{sec:basic} 

In this section, we describe how we estimate the PNS mass and radius from the neutrino signal. First of all, we use a correlation reported in our previous papers \citet{2021MNRAS.500..696N,2021MNRAS.506.1462N}. In both axisymmetric and full 3D CCSN models, we found that the cumulative number of neutrino events in each detector has a strong correlation to the total emitted neutrino energy (TONE)\footnote{More specifically, the TONE is defined as a time-, energy-, and flavor- integrated neutrino radiation in the unit of energy up to a certain post-bounce time.}. The fitting function for the correlation was provided in \citet{2021MNRAS.500..696N}, but it was updated in the subsequent paper \citep{2021MNRAS.506.1462N} (see also Eqs.~\ref{eq:fitSKNORMAL}-\ref{eq:fitIceCubeInV}), which covers the longer post-bounce time (up to $\sim 4$s after bounce) than the former. We also note that the correlation is insensitive to progenitors. One may wonder why each neutrino detector, which usually has a sensitivity on a specific neutrino flavor, is capable of estimating flavor-integrated quantity. There is a physical reason behind the correlation. Let us explain them in the case with SK (HK) as an example. First, the TONE is dominated by $\barparena{\nu}_x$ at the CCSN source except for very early post-bounce phase ($\lesssim 100$ ms), indicating that the accurate estimation of $\barparena{\nu}_x$ at the source is a key.
On the other hand, the $\bar{\nu}_x$s experience flavor conversions during the travel up to the Earth, and some fraction arrive at SK as $\bar{\nu}_e$. They can, hence, be detected through inverse beta decay on protons (IBD-p), the dominant detector channel at SK. This indicates that the observed neutrinos at SK have information on $\bar{\nu}_x$ at CCSN source, which is why the event count has a correlation to TONE. We can interpret the correlation by the similar way for other detectors.

By assuming a neutrino oscillation model\footnote{It should be noted that the uncertainty of neutrino oscillation models may be constrained if neutrinos are observed by multiple neutrino detectors, or GWs are observed. We discuss this in Sec.~\ref{sec:lim} in more detail.}, we can estimate TONE from observed neutrino data at each detector. We also note that TONE is time-dependent, i.e., it represents the total emitted neutrino energy up to a given post-bounce time. This reflects both dynamical features and mass accretion history of CCSN core, indicating that information on PNS structure is expected to be imprinted. We, hence, inspect relations between time evolution of TONE and PNS structure in our multi-D CCSN models. As summarized in Sec.~\ref{sec:correlationTONEPNSM}, we find interesting correlations between them. We fit the relation by polynomial functions, which can be directly used for data analysis in real observation. After briefly summarizing our CCSN models, we describe the detail of our correlation study in Sec.~\ref{sec:correlationTONEPNSM}. We then present how we use our results to real observation in Sec.~\ref{sec:demo}.

\begin{figure}
  \rotatebox{0}{
    \begin{minipage}{1.0\hsize}
        \includegraphics[width=\columnwidth]{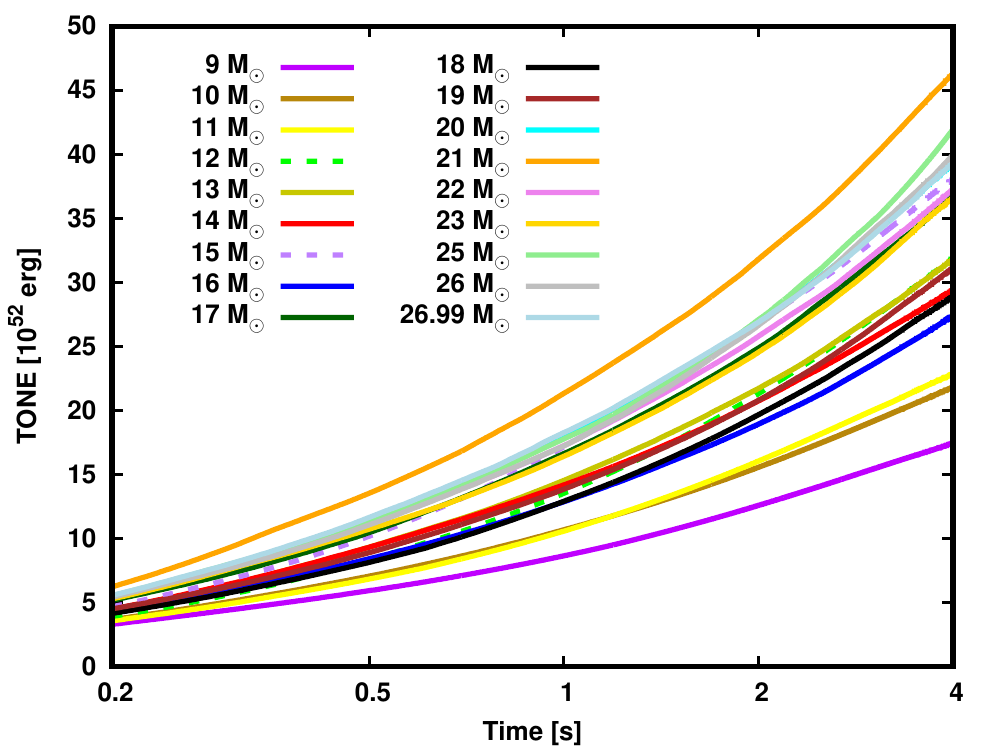}
    \caption{Total emitted neutrino energy (TONE) as a function of post-bounce time. Color distinguishes models. Solid and dashed lines represent explosion and non-explosion models, respectively.
    }
    \label{graph_time_Ecum}
  \end{minipage}}
\end{figure}

\begin{figure*}
  \rotatebox{0}{
    \begin{minipage}{1.0\hsize}
        \includegraphics[width=\columnwidth]{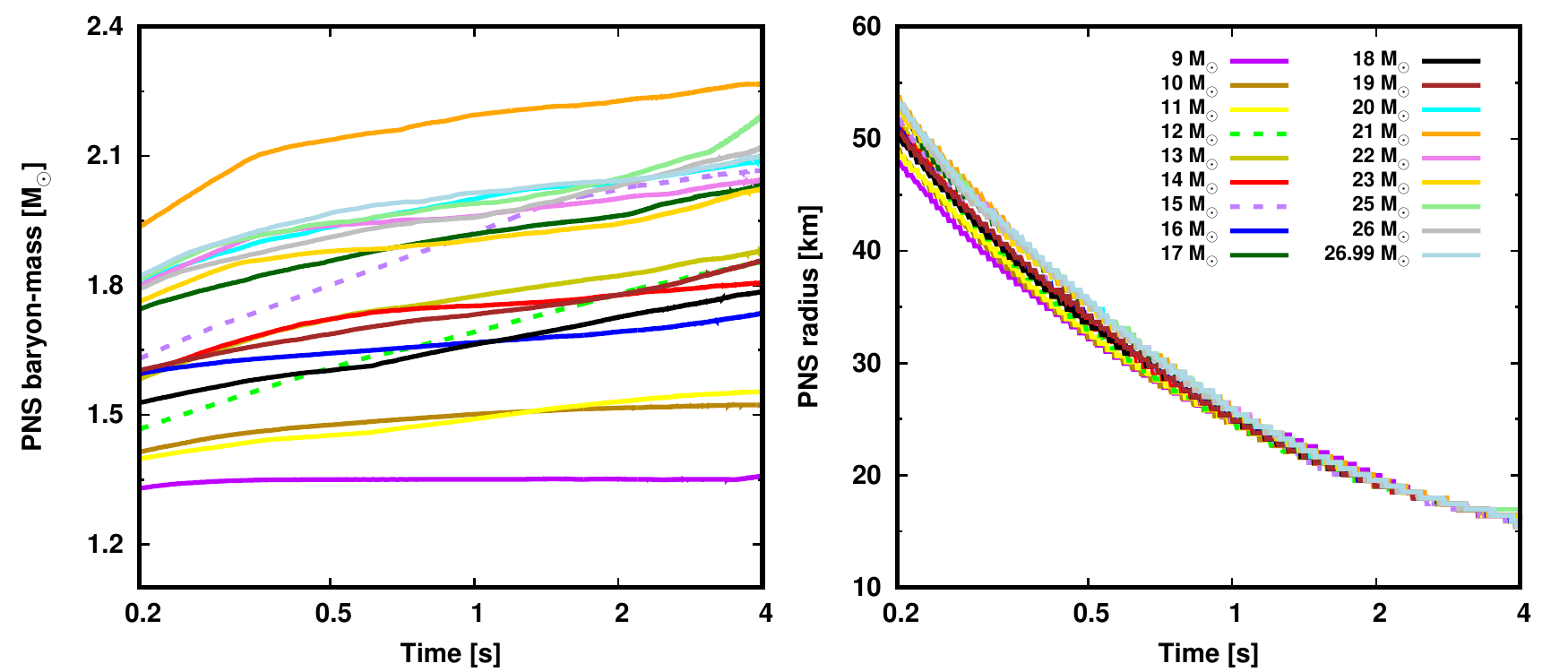}
    \caption{Time evolution of PNS mass (left) and radius (right). In this plot, we determine the PNS radius by the angle-averaged isodensity with $10^{11} {\rm g/cm^3}$.}
    \label{graph_time_PNSM_and_PNSR}
  \end{minipage}}
\end{figure*}

\section{CCSN models}\label{sec:ccsnmodels} 

Before embarking on our analysis, we summarize the multi-D CCSN models employed in this paper. They were simulated by a multi-D neutrino radiation hydrodynamic code, F{\sc{ornax}} \citep{2019ApJS..241....7S}. This incorporates a multi-group neutrino transport scheme based on two-moment approach with up-to-date neutrino-matter interactions following \citep{2002PhRvD..65d3001H,2006NuPhA.777..356B,2017PhRvC..95b5801H} and taking into account the lowest-order corrections of fluid-dependent and general relativistic effects \citep{2006A&A...445..273M}. In this analysis, we employ the most recent axisymmetric (2D) CCSN models reported in \citet{2021Natur.589...29B}, although the 3D models are also available in \citet{2019MNRAS.482..351V,2020MNRAS.491.2715B}. The reason of this choice is that the 2D simulation is computationally much cheaper than that of 3D, which allows the simulation of CCSN for a longer time ($\sim 4$ s after core bounce). We also find that the angular-averaged neutrino signal is almost the same as that obtained from 3D models; hence, we adopt the angle-averaged neutrino data of the 2D models in this study. These models cover the most of accretion phase in CCSN, which is the focused phase in this study.

In these simulations, we cover a wide range of progenitor masses, spanning a zero-age main sequence mass of $9 - 25$~$M_{\sun}$. The initial conditions for the stellar progenitors are provided in \cite{2018ApJ...860...93S}, and 2D simulations were calculated, following the stellar collapse and post-bounce evolution through $\sim 4$ s. Among the (18) models, shock revival is achieved for all except for the 12- and 15~$M_{\sun}$ models. The detailed analysis of their CCSN dynamics can be found in \citep{2021Natur.589...29B}, and that of the neutrino signal are presented in \citep{2021MNRAS.506.1462N}.

Figure~\ref{graph_time_Ecum} shows the time evolution of TONE for our CCSN models. As discussed in \citep{2021MNRAS.506.1462N}, its time evolution has rich progenitor-dependent features. As shown in the figure, the $9 M_{\sun}$ model has the lowest TONE among all models. This model has the steepest density gradient around the core at the presupernova phase \citep[see Fig. 1 in][]{2021Natur.589...29B}, indicating that the mass accretion rate becomes the smallest among our CCSN models. This suppresses the accretion component of neutrino emission, resulting in the lowest TONE. On the contrary, $21 M_{\sun}$ model has the highest TONE. Contrary to the case of $9 M_{\sun}$ model, it has the shallowest density gradient in the core at the presupernova stage, leading the highest mass accretion rate onto PNS and hence the highest TONE.

We show the time evolution of the PNS mass in the left panel of Fig.~\ref{graph_time_PNSM_and_PNSR} obtained from our CCSN simulations. This displays the mass accretion history for each CCSN model, which clearly shows that $9$ and $21 M_{\sun}$ models have the lowest and highest mass accretion rate onto the PNS, respectively (consistent with the above discussion). By comparing TONE and PNS mass, the correlation is obvious; the TONE becomes higher for larger PNS mass. In the next section (Sec.~\ref{sec:correlationTONEPNSM}), we quantify the correlation. In the right panel of Fig.~\ref{graph_time_PNSM_and_PNSR}, on the other hand, we display PNS radius as a function of time. We find that the higher PNS mass tends to have the larger radius, and that the PNS shrinks monotonically with time. It should be noted that the progenitor dependence of PNS radius becomes weaker with increasing time; indeed, all models eventually follow the universal time evolution at $\gtrsim 1$ s. We also quantify these time-dependent features of PNS radius in the next section.

\section{Correlation between TONE and PNS structure}\label{sec:correlationTONEPNSM} 

\begin{figure*}
  \rotatebox{0}{
    \begin{minipage}{1.0\hsize}
        \includegraphics[width=\columnwidth]{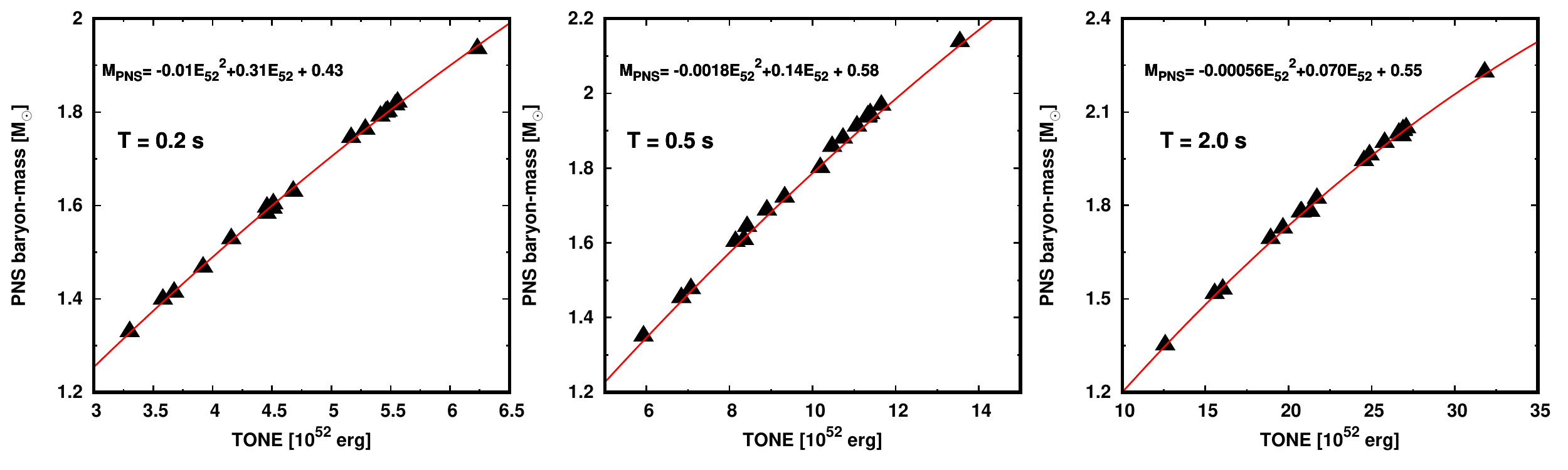}
    \caption{Relation between PNS mass and TONE at a given post-bounce time. From left to right, we display the result at $0.2$ s, $0.5$ s, and $2$ s after bounce. The triangle points represent simulation results, and we fit them with a quadratic function. The coefficients are also displayed in each panel; $M_{\rm PNS}$ and $E_{\rm 52}$ denote the PNS mass (with the unit of $M_{\sun}$) and TONE (with the unit of $10^{52} {\rm erg}$), respectively. }
    \label{graph_Ecum_PNSmass_Cor_fixedT}
  \end{minipage}}
\end{figure*}

Let us first inspect a correlation between TONE and PNS mass in the same time snapshots. In Fig.~\ref{graph_Ecum_PNSmass_Cor_fixedT}, we collect TONE and PNS mass of each CCSN model at the time of $0.2$, $0.5$ and $2$ s in each panel. As illustrated in the plot, the PNS mass has a strong correlation to TONE. The red line in each panel is a quadratic fit for the correlation; the coefficients are displayed in each panel.

\begin{figure}
  \rotatebox{0}{
    \begin{minipage}{1.0\hsize}
        \includegraphics[width=\columnwidth]{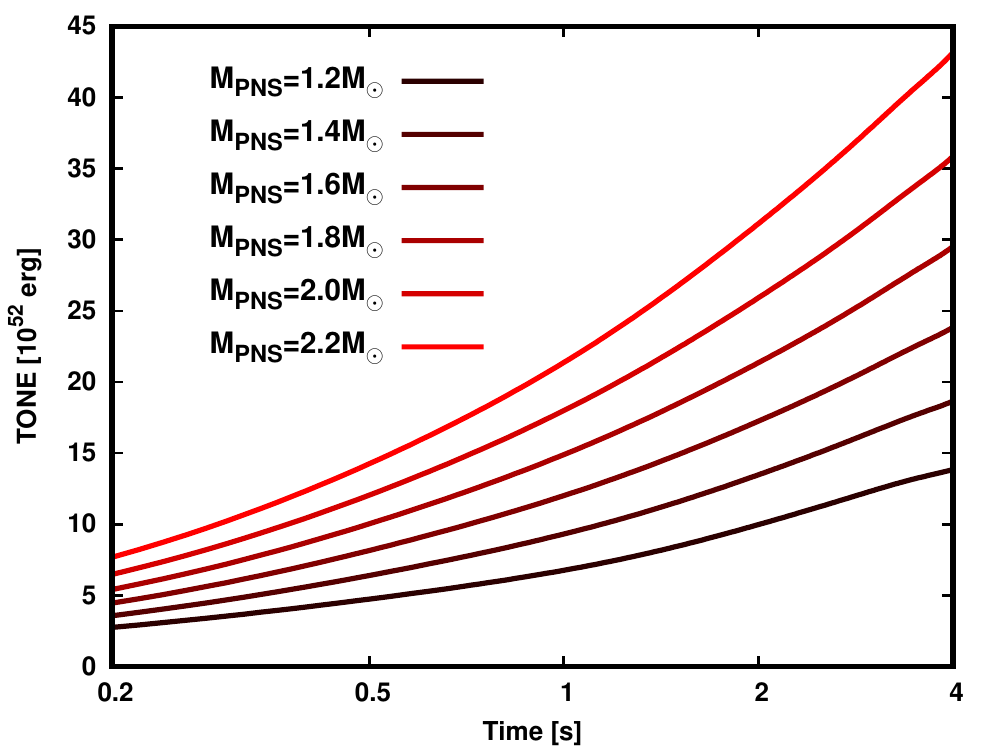}
    \caption{TONE as a function of time along a constant PNS baryon-mass: $1.2, 1.4, 1.6, 1.8, 2.0,$ and $2.2 M_{\sun}$. The fitting function of each line is summarized in Table~\ref{tab:fitcoef_time_PNSM}.
    }
    \label{graph_Ecum_Mconst}
  \end{minipage}}
\end{figure}

It should be mentioned that the fitting function evolves with time, indicating that we can obtain TONE by specifying PNS mass and post-bounce time. In other words, we can draw the time evolution of TONE along a constant PNS mass. We fit them by a seventh degree function;
\begin{eqnarray}
E_{\rm 52}(t) = \sum_{i=0}^{7} a_i t^i,
\label{eq:FitPNSM}
\end{eqnarray}
where $E_{\rm 52}$ denotes TONE in the unit of $10^{52} {\rm erg}$, and $t$ represents the time measured from core bounce in the unit of second. The fitting coefficients for PNS mass in the range of $1.2 - 2.2\, M_{\sun}$ are summarized in Tab.~\ref{tab:fitcoef_time_PNSM}. The time evolution of TONE for selected PNS masses are displayed in Fig.~\ref{graph_Ecum_Mconst}. There are two important remarks in our results. First, our fitting is only valid in the post bounce time of $0.1 \hspace{0.5mm} {\rm s} \lesssim t \lesssim 4 \hspace{0.5mm} {\rm s}$. In the very early post-bounce phase ($\lesssim 0.1$ s), the time evolution of TONE is rather steep, and it would be necessary to use higher polynomials to fit the data. On the other hand, there are other systematic errors in our method at $\lesssim 0.2$ s in our method (see Secs.~\ref{sec:demo}~and~\ref{sec:lim} for more detail); this drawback in our method needs to be improved, although addressing the issue is postponed to future work. We also note that our neutrino data on CCSN models are available up to $\sim 4$ s, indicating that our fitting functions are not reliable after that time. Another remark is that we provide coefficients for PNS mass for each $0.1\, M_{\sun}$ from $1.2 - 2.2\, M_{\sun}$ in Tab.~\ref{tab:fitcoef_time_PNSM}. For cases with other PNS masses, we can simply use a linear interpolation or extrapolation from the adjacent data points.

\begin{figure*}
  \rotatebox{0}{
    \begin{minipage}{1.0\hsize}
        \includegraphics[width=\columnwidth]{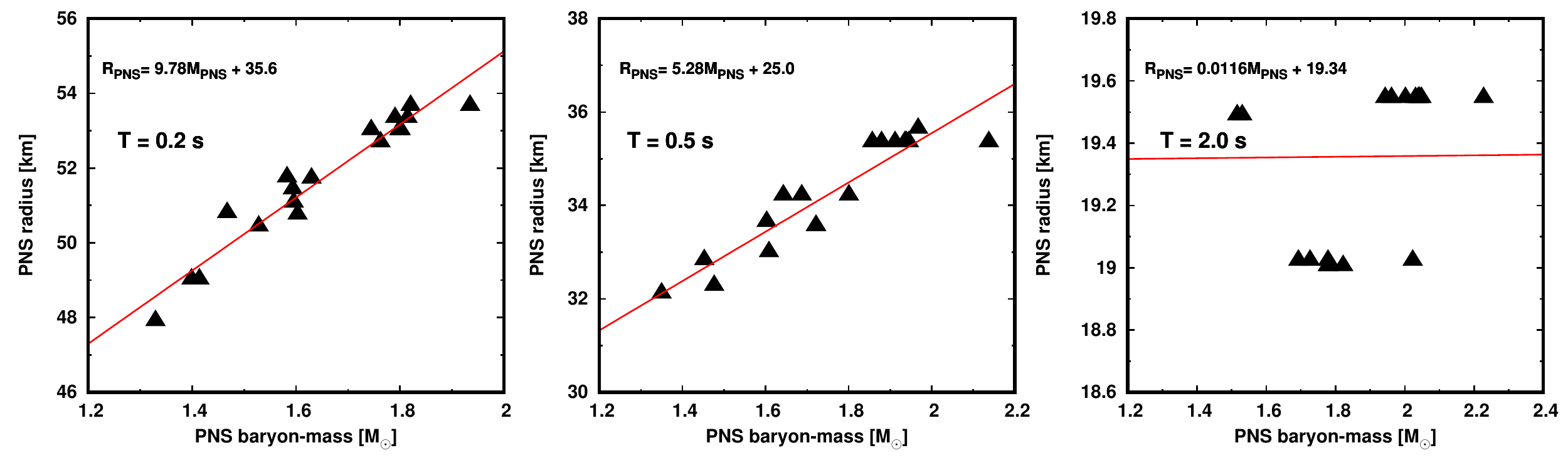}
    \caption{Similar as Fig.~\ref{graph_Ecum_PNSmass_Cor_fixedT} but for relation between PNS mass and radius. We fit the relation with a linear function. The fitting coefficients are displayed in each panel; $M_{\rm PNS}$ and $R_{\rm PNS}$ denote the PNS mass (with the unit of $M_{\sun}$) and radius (with the unit of km), respectively.}
    \label{graph_PNSmass_PNSradi_Cor_fixedT}
  \end{minipage}}
\end{figure*}

As shown in Fig.~\ref{graph_time_PNSM_and_PNSR}, the time evolution of PNS radius is insensitive to CCSN models. However, we find that the PNS radius tends to be (slightly) larger for higher PNS mass at $\lesssim 1$s. We, hence, evaluate the correlation quantitatively; the results are summarized in Fig.~\ref{graph_PNSmass_PNSradi_Cor_fixedT}. As expected, we find that the PNS radius has a positive correlation to its mass at the early post-bounce phase. It should be mentioned that the correlation disappears in the late phase (see right panel of Fig.~\ref{graph_PNSmass_PNSradi_Cor_fixedT}). However, we confirm that the variance of PNS radius is very small (see right panel in Fig.~\ref{graph_PNSmass_PNSradi_Cor_fixedT}); hence, the fitting data is still useful. We also note that the uncertainty of EOS would be more influential to estimate the radius in the late phase, which will be discussed in Sec.~\ref{sec:lim}.

\begin{table*}
\caption{Fitting coefficients for the time evolution of TONE along the constant PNS mass. See Eq.~\ref{eq:FitPNSM} for definition of coefficients.
}
\begin{tabular}{ccccccccc} \hline
~~PNS baryon-mass [$M_{\sun}$]~~ & ~~$a_0$~~ & ~~$a_1$~~ & ~~$a_2$~~ & ~~$a_3$~~ & ~~$a_4$~~ & ~~$a_5$~~ & ~~$a_6$~~ & ~~$a_7$~~ \\
 \hline \hline
1.2 & 0.5333 & 13.93 & -16.66 & 14.34 & -7.168 & 2.039 & -0.3076 & $ 1.909 \times 10^{-2}$ \\
1.3 & 0.5566 & 16.10 & -18.10 & 15.13 & -7.471 & 2.117 & -0.3193 & $ 1.982 \times 10^{-2}$ \\
1.4 & 0.5831 & 18.34 & -19.70 & 16.04 & -7.850 & 2.220 & -0.3348 & $ 2.080 \times 10^{-2}$ \\
1.5 & 0.6135 & 20.66 & -21.43 & 17.11 & -8.318 & 2.351 & -0.3548 & $ 2.207 \times 10^{-2}$ \\
1.6 & 0.6486 & 23.06 & -23.30 & 18.34 & -8.888 & 2.513 & -0.3800 & $ 2.367 \times 10^{-2}$ \\
1.7 & 0.6893 & 25.55 & -25.35 & 19.78 & -9.578 & 2.714 & -0.4113 & $ 2.567 \times 10^{-2}$ \\
1.8 & 0.7371 & 28.15 & -27.59 & 21.47 & -10.41 & 2.959 & -0.4496 & $ 2.813 \times 10^{-2}$ \\
1.9 & 0.7937 & 30.87 & -30.06 & 23.43 & -11.41 & 3.256 & -0.4964 & $ 3.113 \times 10^{-2}$ \\
2.0 & 0.8619 & 33.72 & -32.82 & 25.74 & -12.61 & 3.615 & -0.5530 & $ 3.477 \times 10^{-2}$ \\
2.1 & 0.9456 & 36.72 & -35.90 & 28.47 & -14.06 & 4.048 & -0.6212 & $ 3.916 \times 10^{-2}$ \\
2.2 & 10.508 & 39.89 & -39.38 & 31.69 & -15.78 & 4.567 & -0.7031 & $ 4.443 \times 10^{-2}$ \\
 \hline
\end{tabular}
\label{tab:fitcoef_time_PNSM}
\end{table*}

\begin{table*}
\caption{Fitting coefficients for the time evolution of PNS radius along the constant PNS mass. See Eq.~\ref{eq:FitPNSR} for definition of coefficients.
}
\begin{tabular}{ccccccccc} \hline
~~PNS baryon-mass [$M_{\sun}$]~~ & ~~$b_0$~~ & ~~$b_1$~~ & ~~$b_2$~~ & ~~$b_3$~~ & ~~$b_4$~~ & ~~$b_5$~~ & ~~$b_6$~~ & ~~$b_7$~~ \\
 \hline \hline
1.2 & 2.097 & -3.545 & 4.855 & -3.982 & 1.931 & -0.5429 & $8.163 \times 10^{-2}$ & $-5.062 \times 10^{-3}$ \\
1.3 & 2.140 & -3.689 & 5.146 & -4.302 & 2.120 & -0.6042 & $9.178 \times 10^{-2}$ & $-5.738 \times 10^{-3}$ \\
1.4 & 2.182 & -3.826 & 5.422 & -4.607 & 2.302 & -0.6629 & 0.1015 & $-6.387 \times 10^{-3}$ \\
1.5 & 2.223 & -3.957 & 5.686 & -4.899 & 2.477 & -0.7193 & 0.1109 & $-7.012 \times 10^{-3}$ \\
1.6 & 2.262 & -4.082 & 5.938 & -5.179 & 2.644 & -0.7735 & 0.1199 & $-7.613 \times 10^{-3}$ \\
1.7 & 2.299 & -4.201 & 6.178 & -5.448 & 2.805 & -0.8258 & 0.1286 & $-8.193 \times 10^{-3}$ \\
1.8 & 2.336 & -4.314 & 6.409 & -5.707 & 2.960 & -0.8762 & 0.1369 & $-8.754 \times 10^{-3}$ \\
1.9 & 2.371 & -4.424 & 6.630 & -5.955 & 3.110 & -0.9249 & 0.1450 & $-9.296 \times 10^{-3}$ \\
2.0 & 2.406 & -4.528 & 6.842 & -6.195 & 3.255 & -0.9720 & 0.1529 & $-9.821 \times 10^{-3}$ \\
2.1 & 2.439 & -4.629 & 7.047 & -6.427 & 3.395 & -1.0177 & 0.1605 & $-1.033 \times 10^{-2}$ \\
2.2 & 2.472 & -4.725 & 7.244 & -6.650 & 3.531 & -1.0619 & 0.1679 & $-1.082 \times 10^{-2}$ \\
 \hline
\end{tabular}
\label{tab:fitcoef_time_PNSR}
\end{table*}

\begin{figure}
  \rotatebox{0}{
    \begin{minipage}{1.0\hsize}
        \includegraphics[width=\columnwidth]{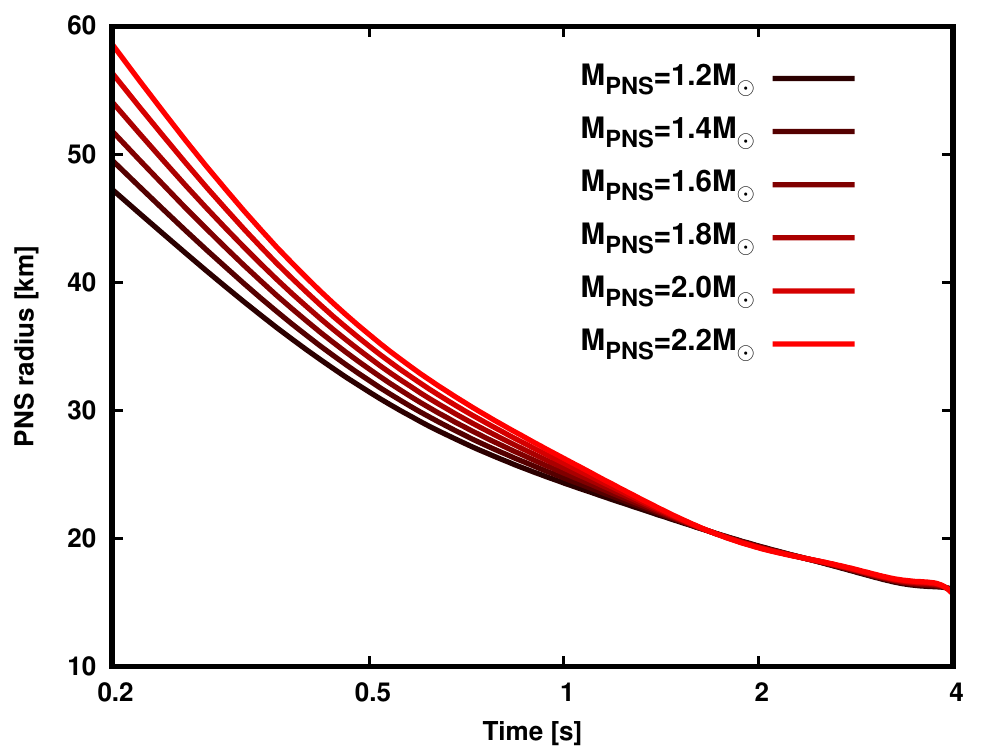}
    \caption{PNS radius as a function of time along a constant PNS mass: $1.2, 1.4, 1.6, 1.8, 2.0,$ and $2.2 M_{\sun}$. In Table~\ref{tab:fitcoef_time_PNSR}, we provide the fitting function of each line.
    }
    \label{graph_RPNS_Mconst}
  \end{minipage}}
\end{figure}

We fit the relation between PNS mass and radius linearly at each time snapshot. This allows us to estimate PNS radius by giving PNS mass and time. This indicates that we can draw the time evolution of PNS radius for a constant PNS mass. We fit them by polynomial functions as
\begin{eqnarray}
\ln R_{\rm 10}(t) = \sum_{i=0}^{7} b_i t^i,
\label{eq:FitPNSR}
\end{eqnarray}
where $R_{\rm 10}$ denotes the PNS radius in the unit of $10$ km; the fitting coefficients are summarized in Tab.~\ref{tab:fitcoef_time_PNSR}. We also draw the time evolution of PNS radius for selected PNS masses in Fig.~\ref{graph_RPNS_Mconst}. In the next section, we demonstrate how these fitting functions can be used for data analysis in real observations.

\section{Demonstration}\label{sec:demo}
In this section, we demonstrate retrievals of time evolution of PNS mass and radius from observed neutrino data by using our proposed method. For the input data, we employ mock data of observed neutrinos in \citet{2021MNRAS.506.1462N}, which were computed by a detector software, SNOwGLoBES\footnote{The software is available at \url{https://webhome.phy.duke.edu/~schol/snowglobes/}.}. The original CCSN models for these mock data are the same as those used in this paper \citep{2021Natur.589...29B}. By assuming neutrino oscillation models and the distance to CCSN, we estimated the energy- and flavor dependent neutrino flux at Earth, and then the neutrino event count at each detector were estimated through SNOwGLoBES (see \citet{2021MNRAS.506.1462N} in more detail). In this study, we consider cases for representative terrestrial neutrino observatories: SK (HK), DUNE, JUNO, and IceCube; and their detector volume is assumed to be $32.5 (220)$ ktons, $40$ ktons, $20$ ktons, and $3.5$ Mtons, respectively. For simplicity, we only consider the major reaction channel at each detector: IBD-p for SK, HK, and IceCube; the charged-current reaction with argon for DUNE. For neutrino oscillation models, we adopt adiabatic Mikheyev-Smirnov-Wolfenstein (MSW) model for both normal- and inverted mass hierarchy. The uncertainty of neutrino oscillation model will be discussed in Sec.~\ref{sec:lim}. In this study, we do not take into account Poisson noise, whereas the smearing effects in detector response that are equipped with SNOwGLoBES are included.

As described in previous sections, we use the time-dependent cumulative number of neutrino events ($N_{\rm Cum}$) at each detector. Under the adiabatic MSW neutrino oscillation model, we can estimate TONE ($E_{52}$) from $N_{\rm Cum}$ as (see also Eqs. 23-30 in \citet{2021MNRAS.506.1462N}),
\begin{eqnarray}
&&\hspace{-16.0mm} {\rm [SK-IBDp-NORMAL]} \nonumber \\
&&\hspace{-13.0mm} N_{\rm Cum} = \left( 220 \hspace{0.5mm} E_{52} + 5 \hspace{0.5mm} E_{52}^2 - 0.074 \hspace{0.5mm} E_{52}^3 + 0.0003 \hspace{0.5mm} E_{52}^4 \right) \nonumber \\
&&\left(  \frac{V}{32.5 \hspace{0.5mm} {\rm ktons}}  \right)
\left(  \frac{d}{10 \hspace{0.5mm} {\rm kpc}}  \right)^{-2}\, ,
\label{eq:fitSKNORMAL} \\
&&\hspace{-16.0mm} {\rm [DUNE-CCAre-NORMAL]} \nonumber \\
&&\hspace{-13.0mm} N_{\rm Cum} = \left( 90 \hspace{0.5mm} E_{52} + 4.5 \hspace{0.5mm} E_{52}^2 - 0.062 \hspace{0.5mm} E_{52}^3 + 0.00028 \hspace{0.5mm} E_{52}^4 \right) \nonumber \\
&&\left(  \frac{V}{40 \hspace{0.5mm} {\rm ktons}}  \right)
\left(  \frac{d}{10 \hspace{0.5mm} {\rm kpc}}  \right)^{-2}\, ,
\label{eq:fitDUNENORMAL} \\
&&\hspace{-16.0mm} {\rm [JUNO-IBDp-NORMAL]} \nonumber \\
&&\hspace{-13.0mm} N_{\rm Cum} = \left( 165 \hspace{0.5mm} E_{52} + 5.1 \hspace{0.5mm} E_{52}^2 - 0.082 \hspace{0.5mm} E_{52}^3 + 0.00039 \hspace{0.5mm} E_{52}^4  \right) \nonumber \\
&&\left(  \frac{V}{20 \hspace{0.5mm} {\rm ktons}}  \right)
\left(  \frac{d}{10 \hspace{0.5mm} {\rm kpc}}  \right)^{-2}\, ,
\label{eq:fitJUNONORMAL} \\
&&\hspace{-16.0mm} {\rm [IceCube-IBDp-NORMAL]} \nonumber \\
&&\hspace{-13.0mm} N_{\rm Cum} = \left( 23000 \hspace{0.5mm} E_{52} + 600 \hspace{0.5mm} E_{52}^2 - 9 \hspace{0.5mm} E_{52}^3 + 0.04 \hspace{0.5mm} E_{52}^4  \right) \nonumber \\
&&\left(  \frac{V}{3.5 \hspace{0.5mm} {\rm Mtons}}  \right)
\left(  \frac{d}{10 \hspace{0.5mm} {\rm kpc}}  \right)^{-2}\, ,
\label{eq:fitIceCubeNORMAL}
\end{eqnarray}
in the normal mass hierarchy; $V$ denotes the detector volume. In the case with the inverted mass hierarchy, the functions can be given as,
\begin{eqnarray}
&&\hspace{-16.0mm} {\rm [SK-IBDp-InV]} \nonumber \\
&&\hspace{-13.0mm}N_{\rm Cum} = \left( 170 \hspace{0.5mm} E_{52} + 4 \hspace{0.5mm} E_{52}^2 - 0.07 \hspace{0.5mm} E_{52}^3 + 0.00036 \hspace{0.5mm} E_{52}^4 \right) \nonumber \\
&&\left(  \frac{V}{32.5 \hspace{0.5mm} {\rm ktons}}  \right)
\left(  \frac{d}{10 \hspace{0.5mm} {\rm kpc}}  \right)^{-2}\, ,
\label{eq:fitSKInV} \\
&&\hspace{-16.0mm} {\rm [DUNE-CCAre-InV]} \nonumber \\
&&\hspace{-13.0mm}N_{\rm Cum} = \left( 90 \hspace{0.5mm} E_{52} + 4.5 \hspace{0.5mm} E_{52}^2 - 0.062 \hspace{0.5mm} E_{52}^3 + 0.00028 \hspace{0.5mm} E_{52}^4 \right) \nonumber \\
&&\left(  \frac{V}{40 \hspace{0.5mm} {\rm ktons}}  \right)
\left(  \frac{d}{10 \hspace{0.5mm} {\rm kpc}}  \right)^{-2}\, ,
\label{eq:fitDUNEInV} \\
&&\hspace{-16.0mm} {\rm [JUNO-IBDp-InV]} \nonumber \\
&&\hspace{-13.0mm}N_{\rm Cum} = \left( 135 \hspace{0.5mm} E_{52} + 3 \hspace{0.5mm} E_{52}^2 - 0.051 \hspace{0.5mm} E_{52}^3 + 0.0003 \hspace{0.5mm} E_{52}^4 \right) \nonumber \\
&&\left(  \frac{V}{20 \hspace{0.5mm} {\rm ktons}}  \right)
\left(  \frac{d}{10 \hspace{0.5mm} {\rm kpc}}  \right)^{-2}\, ,
\label{eq:fitJUNOInV} \\
&&\hspace{-16.0mm} {\rm [IceCube-IBDp-InV]} \nonumber \\
&&\hspace{-13.0mm}N_{\rm Cum} = \left( 18000 \hspace{0.5mm} E_{52} + 430 \hspace{0.5mm} E_{52}^2 - 7 \hspace{0.5mm} E_{52}^3 + 0.035 \hspace{0.5mm} E_{52}^4 \right) \nonumber \\
&&\left(  \frac{V}{3.5 \hspace{0.5mm} {\rm Mtons}}  \right)
\left(  \frac{d}{10 \hspace{0.5mm} {\rm kpc}}  \right)^{-2}\, .
\label{eq:fitIceCubeInV}
\end{eqnarray}
By inserting $V=220 {\rm ktons}$ into Eqs.~\ref{eq:fitSKNORMAL} and \ref{eq:fitSKInV}, they represent the correlation for the case with HK\footnote{Since we do not take into account Poisson noise in this demonstration, both SK and HK provide the identical result. It should be stressed, however, that the reduction of Poisson noise is important to increase the reliability of estimating PNS mass and radius in real observations. See also Sec.~\ref{sec:lim} for the discussion.}.

\begin{figure*}
  \rotatebox{0}{
    \begin{minipage}{1.0\hsize}
        \includegraphics[width=\columnwidth]{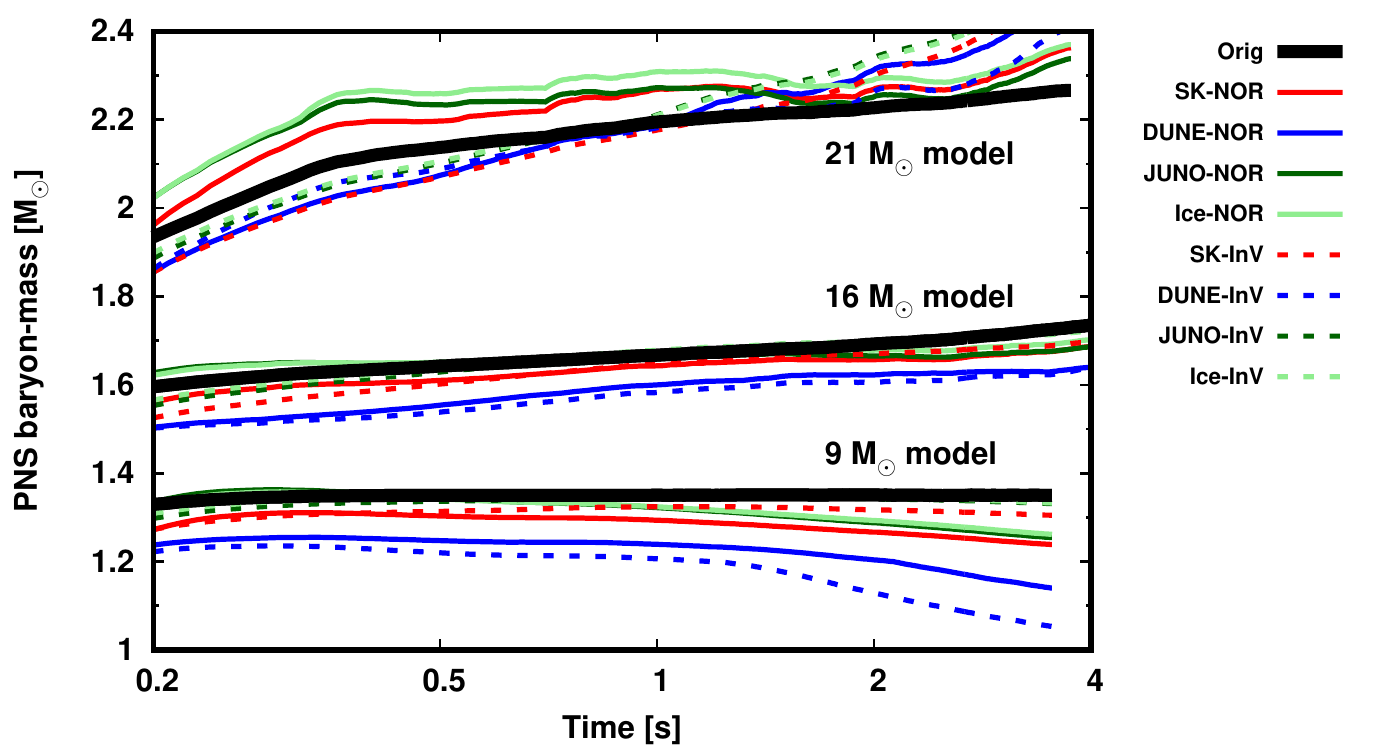}
    \caption{Comparing time evolution of retrieved PNS mass and the original computed from CCSN simulations. We select $9$, $16$, and $21 M_{\sun}$ models as representative examples. The color and line-type distinguish the detector and neutrino oscillation model, respectively.
    }
    \label{graph_tevo_PNSM_reco_selectedModels}
  \end{minipage}}
\end{figure*}

\begin{figure}
  \rotatebox{0}{
    \begin{minipage}{1.0\hsize}
        \includegraphics[width=\columnwidth]{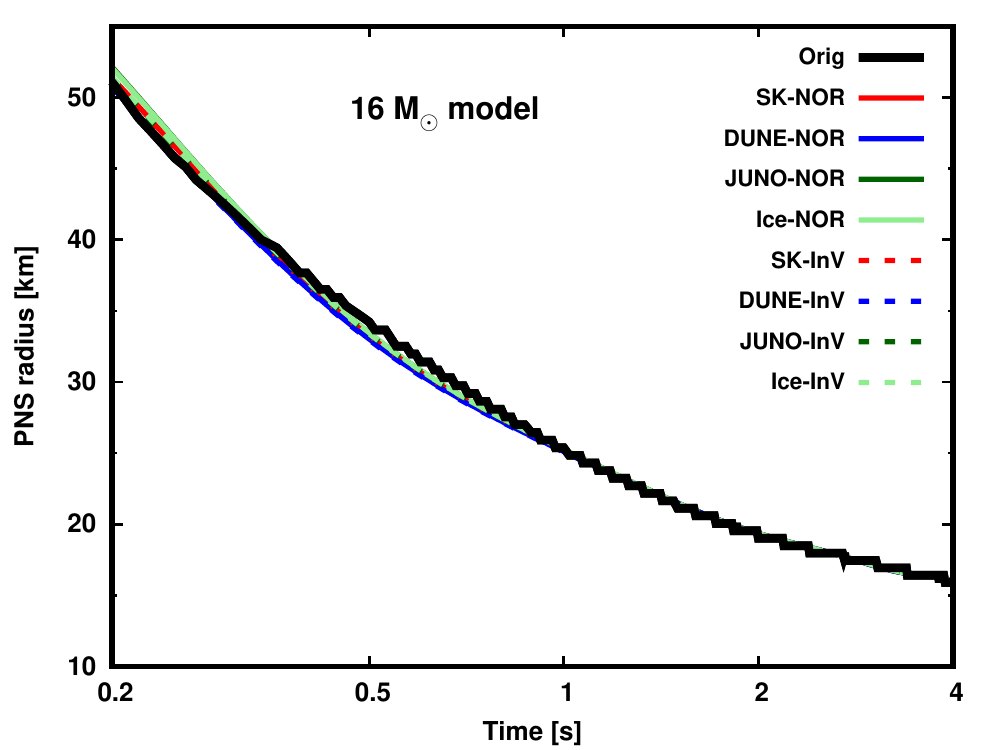}
    \caption{Same as Fig.~\ref{graph_tevo_PNSM_reco_selectedModels} but for PNS radius. We only display the result of $16 M_{\sun}$ CCSN model.
    }
    \label{graph_tevo_PNSR_reco_selectedModels}
  \end{minipage}}
\end{figure}

It should be mentioned that these fitting functions can be applied at all post-bounce time except for the early phase ($\lesssim 0.2$ s)\footnote{We find that there are some large systematic errors in the fitting functions for correlation between $N_{\rm Cum}$ and $E_{52}$ at the early post-bounce phase. This is mainly attributed to the fact that the fitting was made by mainly focusing on the later phase. We postpone the improvement to future work.}; thus, we attempt to retrieve PNS mass and radius by using $N_{\rm Cum}$ of each detector after that time. We can find $E_{52}$ by solving Eqs.~\ref{eq:fitSKNORMAL}~and~\ref{eq:fitSKInV} by using, for instance, Newton's method. From the obtained TONE ($E_{52}$), we can estimate PNS mass by Eq.~\ref{eq:FitPNSM} and PNS radius from Eq.~\ref{eq:FitPNSR} via the obtained PNS mass. In Fig.~\ref{graph_tevo_PNSM_reco_selectedModels}, we compare the retrieved PNS mass to the original in cases with $9$, $16$, and $21 M_{\sun}$ models. We find that the deviation is mainly due to errors in retrievals of $E_{52}$ from $N_{\rm Cum}$. For instance, the retrieved $E_{52}$ in $9 M_{\sun}$ tends to be smaller than the original. We find that the average energy of detected counts at each detector is much smaller than other CCSN models, indicating that the fitting formulae in Eqs.~\ref{eq:fitSKNORMAL}-\ref{eq:fitIceCubeInV} yield smaller $E_{52}$. We are currently improving the fitting formula by using the average energy of event counts. The results will be reported in another paper. Nevertheless, it should be emphasized that the present results are in reasonably agreement with the originals (the error is $\lesssim 15 \%$). In Fig.~\ref{graph_tevo_PNSR_reco_selectedModels}, we compare the retrieved PNS radius to the originals. We find that the error is much smaller than the case with PNS mass, which is due to the insensitiveness of PNS radius to the mass. This indicates that the PNS radius will be estimated accurately if the bounce time can be determined in real observations (but see Sec.~\ref{sec:lim} for the impact of uncertainty of EOS).

\section{Limitations}\label{sec:lim}
Although we have tested the capability of our proposed method in Sec.~\ref{sec:demo}, there are some uncertainties that are not incorporated in the demonstration. In this section, we describe them with discussions.

\begin{enumerate}
\item \underline{Distance to CCSN}:

The estimation of TONE from cumulative number of events at each detector requires the measurement of distance to CCSN. It may be determined from neutrino signal if the neutronization burst at early post-bounce phase (a few milliseconds after core bounce) can be detected. This is by virtue of the so-called Mazurek’s law, leading to universal inner core properties in the collapsing phase \citep[see, e.g.,][]{1985ApJS...58..771B,2003NuPhA.719..144L}. As a result, neutronization burst does not depend on progenitors \citep{2003ApJ...592..434T,2016ApJ...817..182W,2019ApJS..240...38N}, i.e., the burst signal may be used as a standard candle to measure the distance. It should be noted, however, that the detection possibility of neutronization burst depends on neutrino oscillation models, which is one of the major uncertainties in our analysis (see below), indicating that other signals such as electro-magnetic waves and GWs would be necessary to increase the accuracy of the measurement.

\item \underline{Time of core bounce}:

In our method, we need to identify time of core bounce (see Eqs.~\ref{eq:FitPNSM}~and~\ref{eq:FitPNSR}). Similar as the above discussion, the neutronization burst may be the most useful signal to estimate the time. However, the detectability of neutronization burst would depend on neutrino oscillation models as discussed already. It should also be mentioned, on the other hand, that $\bar{\nu}_e$ and $\barparena{\nu}_x$ become abundant a little later after neutronization burst (a few tens of milliseconds), indicating that the first detection of neutrinos should be within the same time scale regardless of neutrino oscillation models\footnote{However, the statistics depends on the distance to CCSN. See below for more details.}. We also note that GW signal may also be useful to estimate the time of core bounce. If the iron core is rapidly rotating, large amount of GWs would be emitted at the core bounce \citep{2004ApJ...600..834O,2004PhRvD..69h4024S,2006A&A...450.1107O,2008A&A...490..231S,2008PhRvD..78f4056D,2011ApJ...743...30T,2014PhRvD..90d4001A,2017PhRvD..95f3019R,2019ApJ...878...13P}. In cases with no- or slow rotation, the prompt convection at $\sim 10$~ms after core bounce would be the first GW signal from CCSN \citep[see, e.g.,][]{2010PThPh.124..331S}. As such, the joint analysis of the neutrino and GW signal plays an important role to increase the accuracy for the determination of time of core bounce.

\item \underline{Detector noise}:
In our demonstration (Sec.~\ref{sec:demo}), we did not take into account Poisson noise when we generated mock data of neutrino observations. This noise reduces the accuracy of retrievals of PNS structure, in particular at the early post-bounce phase (since the cumulative number of event each detector is small). We also note that the PNS mass sensitively depends on TONE at the early phase, indicating that the noise would generate large errors. We need to keep in mind the impact of Poisson noise (and also detector one) when we apply our method in data analysis.

There is a remark, however. In cases with Galactic CCSN, all detectors considered in this paper are capable of detecting more than thousands of neutrinos at $\ge 0.2$ s, indicating that Poisson noise is less than a few percent of the signal. This is by virtue of the fact that our method only requires energy-integrated event counts. We also note that temporal variations of neutrino signal are out of the scope in our method; indeed, cumulative number of events are only required observed data in our method. The energy-integrated cumulative event count is the highest statistics observed quantities, which guarantees that our method can be applied to real observations without suffering from Poisson noise.

\item \underline{Neutrino oscillation}:
When we reconstruct TONE from cumulative number of events, we need to specify a neutrino oscillation model. This is attributed to the fact that the event rate at each detector hinges on the flavor conversion. In our demonstration, we assumed a simple neutrino oscillation model (adiabatic MSW), and the mock data of neutrino signal was also generated by assuming the same oscillation model. In real data analysis, however, we do not know the neutrino oscillation model, indicating that the uncertainty affects directly the estimation of TONE from cumulative number of events. We also note that, since Eqs.~\ref{eq:fitSKNORMAL}-\ref{eq:fitIceCubeInV} are no longer valid for other oscillation models, we need to investigate the similar correlation study between TONE and cumulative number of events.

Nevertheless, our method proposed in this paper would be still informative in real data analysis, since we can carry out a consistency check. By applying Eqs.~\ref{eq:fitSKNORMAL}-\ref{eq:fitIceCubeInV} to observed neutrino data on each detector, time evolution of a PNS mass and radius can be estimated. This estimation is independent for each detector, indicating that the consistency can be tested.

It should also be mentioned that neutrino detections through neutral current reactions have sensitivity to all flavor of neutrinos, although they are subdominant channels in detectors considered in this paper. On the other hand, there are many currently operating or future planed neutrino detectors that utilize nuclear neutral-current interactions (such as coherent elastic neutrino nucleus scattering) \citep{2014PhRvD..89a3011C,2016JCAP...11..017A,2016PhRvD..94j3009L,2017PhRvD..95f5022B,2017APh....89...51A,2020arXiv201213986B,2020PhRvD.102f3001P,2021JCAP...03..043D,2021JCAP...10..064P,2021arXiv211007730A,2021arXiv210908188B}. If TONE can be estimated from these detectors, our fitting formula (Eqs.~\ref{eq:FitPNSM}~and~\ref{eq:FitPNSR}) can be directly used to estimate PNS mass and radius from these observations. This is an interesting possibility and worth to be investigated.

We also note that GW signal would provide another individual constrains on PNS structure \citep{2017PhRvD..96f3005S,2018ApJ...861...10M,2019PhRvL.123e1102T,2020ApJ...898..139W,2021PhRvD.103f3006B,2021arXiv211003131S}. This suggests that the joint analysis would be useful to constrain not only PNS structure but also neutrino flavor conversion in CCSN \citep[see also][]{2020ApJ...898..139W,2021JCAP...11..021H}. Importantly, many theoretical studies suggest that the collective neutrino oscillation commonly occurs in CCSN environments \citep{2019PhRvD.100d3004A,2019ApJ...886..139N,2019ApJ...883...80S,2020PhRvD.101f3001G,2020PhRvD.101b3018D,2020PhRvR...2a2046M,2020PhRvD.101d3016A,2021PhRvD.103f3033A,2021PhRvD.103f3013C,2021PhRvD.104h3025N,2021arXiv211008291H}, indicating that the flavor structure in neutrino signal may be much more complicated than that we expect. Constraining neutrino oscillation models by this proposed method is an interesting possibility.

\item \underline{Angular dependence}:

In our method, we employ angular-averaged neutrino signal obtained from CCSN models. However, the multi-D fluid dynamics in CCSN core generally generates angular-dependent neutrino emission, indicating that the observed neutrino signal depends on the angular location of the observer. The degree of angular variation hinges on progenitors, and the asymmetric degree is $\sim 20 \%$  in our 3D CCSN models \citep{2021MNRAS.500..696N}. In these models, the asymmetric neutrino emission is associated with LESA \citep{2019MNRAS.489.2227V}, which may be a general characteristics in CCSN; indeed, different CCSN groups have reported LESA in their multi-D CCSN models \citep{2014ApJ...792...96T,2018ApJ...865...81O,2019ApJ...881...36G,2019MNRAS.487.1178P}. We also note that PNS kick potentially generates large-scale asymmetric neutrino emission as reported in \citet{2019ApJ...880L..28N}. We need to keep in mind the angular dependence of neutrino signal in real data analysis as an important limitation.

\item \underline{Uncertainties in CCSN models}:

Aside from input physics (see below), there are some uncertainties in our CCSN models. First, we do not take into account stellar rotation, which affects the neutrino luminosity and average energy \citep[see recent studies, e.g.,][]{2018ApJ...852...28S,2021arXiv211100022C}. It may also enhance the degree of angular dependence. The actual impact hinges on the degree of rotation and the rotational profile of progenitor, suggesting that this needs to be investigated systematically. We also note that our CCSN models employ progenitor models computed by spherically symmetric stellar evolution code \citep{2018ApJ...860...93S}. However, multi-D stellar evolution model, which is more realistic than the spherical one, is necessary, since progenitor asymmetries seem to affect both shock revival and neutrino signal \citep[see, e.g.,][]{2013ApJ...778L...7C,2015MNRAS.448.2141M,2017MNRAS.472..491M,2019MNRAS.483..208N,2020MNRAS.493.3496A,2021MNRAS.503.3617A,2021ApJ...908...44Y,2021arXiv210910920V}. We also note that binary stellar evolution models also yield different CCSN dynamics \citep{2015MNRAS.454.3073S,2021ApJ...916L...5V}. It is of importance to gauze the sensitivity of our results to these uncertainties in progenitor models.

Magnetic field is another missing ingredient in our CCSN models. It may alter both CCSN dynamics and neutrino signal, which also hinges on its strength. The stellar rotation would play an important role to amplify the magnetic field, although the detailed investigation of the process is still under investigation \citep[see, e.g.,][as a recent effort]{2021arXiv210813864O}. We leave the task of these systematic studies to future work.

\item \underline{Uncertainties in input physics}:

In this study, we employ CCSN models simulated by F{\sc{ornax}} code. Although this incorporates the up-to-date input physics, there still remain some uncertainties. One of the major uncertainties relevant to the present study may be EOS. In our CCSN simulations, we employ SFHo EOS \citep{2013ApJ...774...17S} that was developed so as to be consistent with both experimental and astrophysical constraints \citep[see, e.g.,][]{2017ApJ...848..105T}. However, some recent results suggest that the slope of symmetry energy may be inconsistent between laboratory experiment and astrophysical constraint \citep[see, e.g.,][]{2021PhRvL.126q2502A,2021PhRvL.126q2503R}, indicating that SFHo EOS may not be a representative theoretical model.

It should be pointed out, however, that the uncertainty of nuclear EOS seems to have a minor influence on our analysis in the post bounce phase at $\lesssim 1$ s. This is attributed to the fact that the thermal effect smears out differences in nuclear properties. On the other hand, the uncertainty would strongly impact estimates of the PNS radius at the late phase. Thus, we need to revise the fitting formula of PNS radius (Eq.~\ref{eq:FitPNSR}), if the nuclear parameters in SFHo EOS does not represent the realistic property.

Uncertainties of nuclear weak interactions at high matter density also give an impact on estimation of PNS structure from neutrino signal. $\nu_x$ emission may be the most affected among neutrino species, since the neutrino sphere would be located at the inner most region among all neutrino species. The ambiguity smears out the relations between PNS mass and TONE, and it also affects the correlation between TONE and cumulative number of events at each detector.

As other uncertainties, we employ approximate treatments in neutrino transport and relativistic corrections for F{\sc{ornax}} CCSN models. This should be improved by general relativistic CCSN simulations with full Boltzmann neutrino transport, and the project towards the ultimate CCSN model is on going \citep[see, e.g.,][]{2018ApJ...854..136N,2019ApJ...880L..28N,2019ApJ...872..181H,2020ApJ...902..150H,2020ApJ...903...82I,2021arXiv210905846I,2021ApJ...909..210A}. These sustained efforts are important to increase the accuracy of theoretical CCSN models, which will revise the fitting formulae and represent PNS structures more realistically.
\end{enumerate}


\section{Summary}\label{sec:sum}
In this paper, we present a new method to estimate the PNS mass and radius from the neutrino signal. Our method is developed by correlation studies based on state-of-the-art multi-D CCSN models. This is the first attempt to tackle the issue: estimating time evolution of PNS from neutrino signal during the development of CCSN explosion phase taking into account multi-D effects. Despite the fact that the inner dynamics of multi-D CCSN models are very complex, we find some interesting correlations representing the essential and intrinsic relation between neutrino signal and PNS structure. We fit them with simple polynomial functions so as to be convenient for real data analysis. The fitting functions and coefficients are provided in this paper; Eqs.~\ref{eq:fitSKNORMAL}-\ref{eq:fitIceCubeInV}, are used to estimate TONE from time-dependent cumulative number of events at each detector; Eq~\ref{eq:FitPNSM} with coefficients summarized in Tab.~\ref{tab:fitcoef_time_PNSM} allows us to estimate PNS mass from TONE; the PNS radius can be estimated from the obtained PNS mass and time from Eq~\ref{eq:FitPNSR}. We test the capability of our proposed method by demonstrating retrievals of PNS mass and radius from mock data of neutrino observation. The retrieved PNS mass and radius are in reasonably good agreement with the simulated values. We finally describe the limitations in our proposed method, which should be taken into account in real data analysis.

This study paves the way to extract physical information on CCSN and PNS structure from the neutrino signal. Joint analyses with multiple neutrino observatories and with GW detection would improve the accuracy of PNS structure estimates (see Sec.~\ref{sec:lim}). It is also important to note that the time evolution of the PNS mass illustrates the mass accretion rate onto the PNS, which reflects the progenitor structure (e.g., core compactness, \citealt{2013ApJ...762..126O,2017JPhG...44k4001H}) and also CCSN dynamics. It would be very interesting if we detect the sudden drop of mass accretion rate, since it may be a good indicator of either shock revival or the arrival of an Si/O interface at the post-shock region. However, the detailed investigation is required to assess the feasibility of the analysis strongly hinges on the sensitivity of each detector and distance to CCSN source. As such, our demonstration presented in this paper is just the first step, and there remains work needed to improve the method. We leave these tasks to future work.

\section*{Acknowledgements}
HN acknowledges Kate Scholberg for help using the SNOwGLoBES software, and Eve Armstrong for useful comments and discussions. We are also grateful for ongoing contributions to the effort of CCSN simulation projects by Adam Burrows, David Radice, Josh Dolence, Aaron Skinner, Matthew Coleman, and Chris White. We acknowledge support from the U.S. Department of Energy Office of Science and the Office of Advanced Scientific Computing Research via the Scientific Discovery through Advanced Computing (SciDAC4) program and Grant DE-SC0018297 (subaward 00009650). In addition, we gratefully acknowledge support from the U.S. NSF under Grants AST-1714267 and PHY-1804048 (the latter via the Max-Planck/Princeton Center (MPPC) for Plasma Physics). An award of computer time was provided  by the INCITE program. That research used resources of the Argonne Leadership Computing Facility, which is a DOE Office of Science User Facility supported under Contract DE-AC02-06CH11357. In addition, this overall research project is part of the Blue Waters sustained-petascale computing project, which is supported by the National Science Foundation (awards OCI-0725070 and ACI-1238993) and the state of Illinois. Blue Waters is a joint effort of the University of Illinois at Urbana-Champaign and its National Center for Supercomputing Applications. This general project is also part of the ``Three-Dimensional Simulations of Core-Collapse Supernovae" PRAC allocation support by the National Science Foundation (under award \#OAC-1809073). Moreover, access under the local award \#TG-AST170045 to the resource Stampede2 in the Extreme Science and Engineering Discovery Environment (XSEDE), which is supported by National Science Foundation grant number ACI-1548562, was crucial to the completion of this work. Finally, the authors employed computational resources provided by the TIGRESS high performance computer center at Princeton University, which is jointly supported by the Princeton Institute for Computational Science and Engineering (PICSciE) and the Princeton University Office of Information Technology, and acknowledge our continuing allocation at the National Energy Research Scientific Computing Center (NERSC), which is supported by the Office of Science of the US Department of Energy (DOE) under contract DE-AC03-76SF00098.

\section*{DATA AVAILABILITY}
The data underlying this article will be shared on reasonable request to the corresponding author.





\bibliographystyle{mnras}
\bibliography{bibfile}

\begin{thebibliography}{}
\makeatletter
\relax
\def\mn@urlcharsother{\let\do\@makeother \do\$\do\&\do\#\do\^\do\_\do\%\do\~}
\def\mn@doi{\begingroup\mn@urlcharsother \@ifnextchar [ {\mn@doi@}
  {\mn@doi@[]}}
\def\mn@doi@[#1]#2{\def\@tempa{#1}\ifx\@tempa\@empty \href
  {http://dx.doi.org/#2} {doi:#2}\else \href {http://dx.doi.org/#2} {#1}\fi
  \endgroup}
\def\mn@eprint#1#2{\mn@eprint@#1:#2::\@nil}
\def\mn@eprint@arXiv#1{\href {http://arxiv.org/abs/#1} {{\tt arXiv:#1}}}
\def\mn@eprint@dblp#1{\href {http://dblp.uni-trier.de/rec/bibtex/#1.xml}
  {dblp:#1}}
\def\mn@eprint@#1:#2:#3:#4\@nil{\def\@tempa {#1}\def\@tempb {#2}\def\@tempc
  {#3}\ifx \@tempc \@empty \let \@tempc \@tempb \let \@tempb \@tempa \fi \ifx
  \@tempb \@empty \def\@tempb {arXiv}\fi \@ifundefined
  {mn@eprint@\@tempb}{\@tempb:\@tempc}{\expandafter \expandafter \csname
  mn@eprint@\@tempb\endcsname \expandafter{\@tempc}}}

\bibitem[\protect\citeauthoryear{{Aalbers} et~al.,}{{Aalbers}
  et~al.}{2016}]{2016JCAP...11..017A}
{Aalbers} J.,  et~al., 2016, \mn@doi [\jcap] {10.1088/1475-7516/2016/11/017},
  2016, 017

\bibitem[\protect\citeauthoryear{{Abbar}, {Duan}, {Sumiyoshi}, {Takiwaki}  \&
  {Volpe}}{{Abbar} et~al.}{2019}]{2019PhRvD.100d3004A}
{Abbar} S.,  {Duan} H.,  {Sumiyoshi} K.,  {Takiwaki} T.,   {Volpe} M.~C.,
  2019, \mn@doi [\prd] {10.1103/PhysRevD.100.043004}, 100, 043004

\bibitem[\protect\citeauthoryear{{Abbar}, {Duan}, {Sumiyoshi}, {Takiwaki}  \&
  {Volpe}}{{Abbar} et~al.}{2020}]{2020PhRvD.101d3016A}
{Abbar} S.,  {Duan} H.,  {Sumiyoshi} K.,  {Takiwaki} T.,   {Volpe} M.~C.,
  2020, \mn@doi [\prd] {10.1103/PhysRevD.101.043016}, 101, 043016

\bibitem[\protect\citeauthoryear{{Abbar}, {Capozzi}, {Glas}, {Janka}  \&
  {Tamborra}}{{Abbar} et~al.}{2021}]{2021PhRvD.103f3033A}
{Abbar} S.,  {Capozzi} F.,  {Glas} R.,  {Janka} H.~T.,   {Tamborra} I.,  2021,
  \mn@doi [\prd] {10.1103/PhysRevD.103.063033}, 103, 063033

\bibitem[\protect\citeauthoryear{{Abbasi} et~al.,}{{Abbasi}
  et~al.}{2011}]{2011A&A...535A.109A}
{Abbasi} R.,  et~al., 2011, \mn@doi [\aap] {10.1051/0004-6361/201117810}, 535,
  A109

\bibitem[\protect\citeauthoryear{{Abdikamalov} \& {Foglizzo}}{{Abdikamalov} \&
  {Foglizzo}}{2020}]{2020MNRAS.493.3496A}
{Abdikamalov} E.,  {Foglizzo} T.,  2020, \mn@doi [\mnras]
  {10.1093/mnras/staa533}, 493, 3496

\bibitem[\protect\citeauthoryear{{Abdikamalov}, {Gossan}, {DeMaio}  \&
  {Ott}}{{Abdikamalov} et~al.}{2014}]{2014PhRvD..90d4001A}
{Abdikamalov} E.,  {Gossan} S.,  {DeMaio} A.~M.,   {Ott} C.~D.,  2014, \mn@doi
  [\prd] {10.1103/PhysRevD.90.044001}, 90, 044001

\bibitem[\protect\citeauthoryear{{Abdikamalov}, {Foglizzo}  \&
  {Mukazhanov}}{{Abdikamalov} et~al.}{2021}]{2021MNRAS.503.3617A}
{Abdikamalov} E.,  {Foglizzo} T.,   {Mukazhanov} O.,  2021, \mn@doi [\mnras]
  {10.1093/mnras/stab715}, 503, 3617

\bibitem[\protect\citeauthoryear{{Abe} et~al.,}{{Abe}
  et~al.}{2016}]{2016APh....81...39A}
{Abe} K.,  et~al., 2016, \mn@doi [Astroparticle Physics]
  {10.1016/j.astropartphys.2016.04.003}, 81, 39

\bibitem[\protect\citeauthoryear{{Abe} et~al.,}{{Abe}
  et~al.}{2017}]{2017APh....89...51A}
{Abe} K.,  et~al., 2017, \mn@doi [Astroparticle Physics]
  {10.1016/j.astropartphys.2017.01.006}, 89, 51

\bibitem[\protect\citeauthoryear{{Abi} et~al.,}{{Abi}
  et~al.}{2020}]{2020arXiv200806647A}
{Abi} B.,  et~al., 2020, arXiv e-prints, p. arXiv:2008.06647

\bibitem[\protect\citeauthoryear{{Acciarri} et~al.,}{{Acciarri}
  et~al.}{2016}]{2016arXiv160105471A}
{Acciarri} R.,  et~al., 2016, preprint (\mn@eprint {arXiv} {1601.05471})

\bibitem[\protect\citeauthoryear{{Adhikari} et~al.,}{{Adhikari}
  et~al.}{2021}]{2021PhRvL.126q2502A}
{Adhikari} D.,  et~al., 2021, \mn@doi [\prl] {10.1103/PhysRevLett.126.172502},
  126, 172502

\bibitem[\protect\citeauthoryear{{Akaho} et~al.,}{{Akaho}
  et~al.}{2021}]{2021ApJ...909..210A}
{Akaho} R.,  et~al., 2021, \mn@doi [\apj] {10.3847/1538-4357/abe1bf}, 909, 210

\bibitem[\protect\citeauthoryear{{Akimov} et~al.,}{{Akimov}
  et~al.}{2021}]{2021arXiv211007730A}
{Akimov} D.,  et~al., 2021, arXiv e-prints, p. arXiv:2110.07730

\bibitem[\protect\citeauthoryear{{An} et~al.,}{{An}
  et~al.}{2016}]{2016JPhG...43c0401A}
{An} F.,  et~al., 2016, \mn@doi [Journal of Physics G Nuclear Physics]
  {10.1088/0954-3899/43/3/030401}, 43, 030401

\bibitem[\protect\citeauthoryear{{Ankowski} et~al.,}{{Ankowski}
  et~al.}{2016}]{2016arXiv160807853A}
{Ankowski} A.,  et~al., 2016, arXiv e-prints, p. arXiv:1608.07853

\bibitem[\protect\citeauthoryear{{Bandyopadhyay}, {Bhattacharjee},
  {Chakraborty}, {Kar}  \& {Saha}}{{Bandyopadhyay}
  et~al.}{2017}]{2017PhRvD..95f5022B}
{Bandyopadhyay} A.,  {Bhattacharjee} P.,  {Chakraborty} S.,  {Kar} K.,   {Saha}
  S.,  2017, \mn@doi [\prd] {10.1103/PhysRevD.95.065022}, 95, 065022

\bibitem[\protect\citeauthoryear{{Baxter} et~al.,}{{Baxter}
  et~al.}{2021}]{2021arXiv210908188B}
{Baxter} A.~L.,  et~al., 2021, arXiv e-prints, p. arXiv:2109.08188

\bibitem[\protect\citeauthoryear{{Bhattacharjee}, {Bandyopadhyay},
  {Chakraborty}, {Ghosh}, {Kar}  \& {Saha}}{{Bhattacharjee}
  et~al.}{2020}]{2020arXiv201213986B}
{Bhattacharjee} P.,  {Bandyopadhyay} A.,  {Chakraborty} S.,  {Ghosh} S.,  {Kar}
  K.,   {Saha} S.,  2020, arXiv e-prints, p. arXiv:2012.13986

\bibitem[\protect\citeauthoryear{{Bionta} et~al.,}{{Bionta}
  et~al.}{1987}]{1987PhRvL..58.1494B}
{Bionta} R.~M.,  et~al., 1987, \mn@doi [\prl] {10.1103/PhysRevLett.58.1494},
  58, 1494

\bibitem[\protect\citeauthoryear{{Bizouard}, {Maturana-Russel},
  {Torres-Forn{\'e}}, {Obergaulinger}, {Cerd{\'a}-Dur{\'a}n}, {Christensen},
  {Font}  \& {Meyer}}{{Bizouard} et~al.}{2021}]{2021PhRvD.103f3006B}
{Bizouard} M.-A.,  {Maturana-Russel} P.,  {Torres-Forn{\'e}} A.,
  {Obergaulinger} M.,  {Cerd{\'a}-Dur{\'a}n} P.,  {Christensen} N.,  {Font}
  J.~A.,   {Meyer} R.,  2021, \mn@doi [\prd] {10.1103/PhysRevD.103.063006},
  103, 063006

\bibitem[\protect\citeauthoryear{{Bruenn}}{{Bruenn}}{1985}]{1985ApJS...58..771B}
{Bruenn} S.~W.,  1985, \mn@doi [\apjs] {10.1086/191056}, 58, 771

\bibitem[\protect\citeauthoryear{{Burrows} \& {Lattimer}}{{Burrows} \&
  {Lattimer}}{1986}]{1986ApJ...307..178B}
{Burrows} A.,  {Lattimer} J.~M.,  1986, \mn@doi [\apj] {10.1086/164405}, 307,
  178

\bibitem[\protect\citeauthoryear{{Burrows} \& {Vartanyan}}{{Burrows} \&
  {Vartanyan}}{2021}]{2021Natur.589...29B}
{Burrows} A.,  {Vartanyan} D.,  2021, \mn@doi [\nat]
  {10.1038/s41586-020-03059-w}, 589, 29

\bibitem[\protect\citeauthoryear{{Burrows}, {Reddy}  \& {Thompson}}{{Burrows}
  et~al.}{2006}]{2006NuPhA.777..356B}
{Burrows} A.,  {Reddy} S.,   {Thompson} T.~A.,  2006, \mn@doi [Nuclear Physics
  A] {10.1016/j.nuclphysa.2004.06.012}, 777, 356

\bibitem[\protect\citeauthoryear{{Burrows}, {Radice}, {Vartanyan}, {Nagakura},
  {Skinner}  \& {Dolence}}{{Burrows} et~al.}{2020}]{2020MNRAS.491.2715B}
{Burrows} A.,  {Radice} D.,  {Vartanyan} D.,  {Nagakura} H.,  {Skinner} M.~A.,
   {Dolence} J.~C.,  2020, \mn@doi [\mnras] {10.1093/mnras/stz3223}, 491, 2715

\bibitem[\protect\citeauthoryear{{Capozzi}, {Abbar}, {Bollig}  \&
  {Janka}}{{Capozzi} et~al.}{2021}]{2021PhRvD.103f3013C}
{Capozzi} F.,  {Abbar} S.,  {Bollig} R.,   {Janka} H.~T.,  2021, \mn@doi [\prd]
  {10.1103/PhysRevD.103.063013}, 103, 063013

\bibitem[\protect\citeauthoryear{{Chakraborty}, {Bhattacharjee}  \&
  {Kar}}{{Chakraborty} et~al.}{2014}]{2014PhRvD..89a3011C}
{Chakraborty} S.,  {Bhattacharjee} P.,   {Kar} K.,  2014, \mn@doi [\prd]
  {10.1103/PhysRevD.89.013011}, 89, 013011

\bibitem[\protect\citeauthoryear{{Chan}, {M{\"u}ller}, {Heger}, {Pakmor}  \&
  {Springel}}{{Chan} et~al.}{2018}]{2018ApJ...852L..19C}
{Chan} C.,  {M{\"u}ller} B.,  {Heger} A.,  {Pakmor} R.,   {Springel} V.,  2018,
  \mn@doi [\apjl] {10.3847/2041-8213/aaa28c}, 852, L19

\bibitem[\protect\citeauthoryear{{Chan}, {M{\"u}ller}  \& {Heger}}{{Chan}
  et~al.}{2020}]{2020MNRAS.495.3751C}
{Chan} C.,  {M{\"u}ller} B.,   {Heger} A.,  2020, \mn@doi [\mnras]
  {10.1093/mnras/staa1431}, 495, 3751

\bibitem[\protect\citeauthoryear{{Coleman}, {Burrows}  \& {White}}{{Coleman}
  et~al.}{2021}]{2021arXiv211100022C}
{Coleman} M. S.~B.,  {Burrows} A.,   {White} C.~J.,  2021, arXiv e-prints, p.
  arXiv:2111.00022

\bibitem[\protect\citeauthoryear{{Couch} \& {Ott}}{{Couch} \&
  {Ott}}{2013}]{2013ApJ...778L...7C}
{Couch} S.~M.,  {Ott} C.~D.,  2013, \mn@doi [The Astrophysical Journal]
  {10.1088/2041-8205/778/1/L7}, 778, L7

\bibitem[\protect\citeauthoryear{{DarkSide-20k Collaboration}
  et~al.,}{{DarkSide-20k Collaboration} et~al.}{2021}]{2021JCAP...03..043D}
{DarkSide-20k Collaboration} et~al., 2021, \mn@doi [\jcap]
  {10.1088/1475-7516/2021/03/043}, 2021, 043

\bibitem[\protect\citeauthoryear{{Delfan Azari} et~al.,}{{Delfan Azari}
  et~al.}{2020}]{2020PhRvD.101b3018D}
{Delfan Azari} M.,  et~al., 2020, \mn@doi [\prd] {10.1103/PhysRevD.101.023018},
  101, 023018

\bibitem[\protect\citeauthoryear{{Dessart}, {Burrows}, {Livne}  \&
  {Ott}}{{Dessart} et~al.}{2006}]{2006ApJ...645..534D}
{Dessart} L.,  {Burrows} A.,  {Livne} E.,   {Ott} C.~D.,  2006, \mn@doi [\apj]
  {10.1086/504068}, 645, 534

\bibitem[\protect\citeauthoryear{{Dimmelmeier}, {Ott}, {Marek}  \&
  {Janka}}{{Dimmelmeier} et~al.}{2008}]{2008PhRvD..78f4056D}
{Dimmelmeier} H.,  {Ott} C.~D.,  {Marek} A.,   {Janka} H.~T.,  2008, \mn@doi
  [\prd] {10.1103/PhysRevD.78.064056}, 78, 064056

\bibitem[\protect\citeauthoryear{{Fischer}, {Guo}, {Dzhioev},
  {Mart{\'\i}nez-Pinedo}, {Wu}, {Lohs}  \& {Qian}}{{Fischer}
  et~al.}{2020}]{2020PhRvC.101b5804F}
{Fischer} T.,  {Guo} G.,  {Dzhioev} A.~A.,  {Mart{\'\i}nez-Pinedo} G.,  {Wu}
  M.-R.,  {Lohs} A.,   {Qian} Y.-Z.,  2020, \mn@doi [\prc]
  {10.1103/PhysRevC.101.025804}, 101, 025804

\bibitem[\protect\citeauthoryear{{Fryer}}{{Fryer}}{2009}]{2009ApJ...699..409F}
{Fryer} C.~L.,  2009, \mn@doi [\apj] {10.1088/0004-637X/699/1/409}, 699, 409

\bibitem[\protect\citeauthoryear{{Gallo Rosso}, {Vissani}  \& {Volpe}}{{Gallo
  Rosso} et~al.}{2017}]{2017JCAP...11..036G}
{Gallo Rosso} A.,  {Vissani} F.,   {Volpe} M.~C.,  2017, \mn@doi [\jcap]
  {10.1088/1475-7516/2017/11/036}, 2017, 036

\bibitem[\protect\citeauthoryear{{Gallo Rosso}, {Abbar}, {Vissani}  \&
  {Volpe}}{{Gallo Rosso} et~al.}{2018}]{2018JCAP...12..006G}
{Gallo Rosso} A.,  {Abbar} S.,  {Vissani} F.,   {Volpe} M.~C.,  2018, \mn@doi
  [\jcap] {10.1088/1475-7516/2018/12/006}, 2018, 006

\bibitem[\protect\citeauthoryear{{Glas}, {Janka}, {Melson}, {Stockinger}  \&
  {Just}}{{Glas} et~al.}{2019}]{2019ApJ...881...36G}
{Glas} R.,  {Janka} H.~T.,  {Melson} T.,  {Stockinger} G.,   {Just} O.,  2019,
  \mn@doi [\apj] {10.3847/1538-4357/ab275c}, 881, 36

\bibitem[\protect\citeauthoryear{{Glas}, {Janka}, {Capozzi}, {Sen}, {Dasgupta},
  {Mirizzi}  \& {Sigl}}{{Glas} et~al.}{2020}]{2020PhRvD.101f3001G}
{Glas} R.,  {Janka} H.~T.,  {Capozzi} F.,  {Sen} M.,  {Dasgupta} B.,  {Mirizzi}
  A.,   {Sigl} G.,  2020, \mn@doi [\prd] {10.1103/PhysRevD.101.063001}, 101,
  063001

\bibitem[\protect\citeauthoryear{{Halim}, {Casentini}, {Drago}, {Fafone},
  {Scholberg}, {Vigorito}  \& {Pagliaroli}}{{Halim}
  et~al.}{2021}]{2021JCAP...11..021H}
{Halim} O.,  {Casentini} C.,  {Drago} M.,  {Fafone} V.,  {Scholberg} K.,
  {Vigorito} C.~F.,   {Pagliaroli} G.,  2021, \mn@doi [\jcap]
  {10.1088/1475-7516/2021/11/021}, 2021, 021

\bibitem[\protect\citeauthoryear{{Harada} \& {Nagakura}}{{Harada} \&
  {Nagakura}}{2021}]{2021arXiv211008291H}
{Harada} A.,  {Nagakura} H.,  2021, arXiv e-prints, p. arXiv:2110.08291

\bibitem[\protect\citeauthoryear{{Harada}, {Nagakura}, {Iwakami}, {Okawa},
  {Furusawa}, {Matsufuru}, {Sumiyoshi}  \& {Yamada}}{{Harada}
  et~al.}{2019}]{2019ApJ...872..181H}
{Harada} A.,  {Nagakura} H.,  {Iwakami} W.,  {Okawa} H.,  {Furusawa} S.,
  {Matsufuru} H.,  {Sumiyoshi} K.,   {Yamada} S.,  2019, \mn@doi [\apj]
  {10.3847/1538-4357/ab0203}, 872, 181

\bibitem[\protect\citeauthoryear{{Harada}, {Nagakura}, {Iwakami}, {Okawa},
  {Furusawa}, {Sumiyoshi}, {Matsufuru}  \& {Yamada}}{{Harada}
  et~al.}{2020}]{2020ApJ...902..150H}
{Harada} A.,  {Nagakura} H.,  {Iwakami} W.,  {Okawa} H.,  {Furusawa} S.,
  {Sumiyoshi} K.,  {Matsufuru} H.,   {Yamada} S.,  2020, \mn@doi [\apj]
  {10.3847/1538-4357/abb5a9}, 902, 150

\bibitem[\protect\citeauthoryear{{Hirata} et~al.,}{{Hirata}
  et~al.}{1987}]{1987PhRvL..58.1490H}
{Hirata} K.,  et~al., 1987, \mn@doi [\prl] {10.1103/PhysRevLett.58.1490}, 58,
  1490

\bibitem[\protect\citeauthoryear{{Horiuchi} \& {Kneller}}{{Horiuchi} \&
  {Kneller}}{2018}]{2018JPhG...45d3002H}
{Horiuchi} S.,  {Kneller} J.~P.,  2018, \mn@doi [Journal of Physics G Nuclear
  Physics] {10.1088/1361-6471/aaa90a}, 45, 043002

\bibitem[\protect\citeauthoryear{{Horiuchi}, {Nakamura}, {Takiwaki}  \&
  {Kotake}}{{Horiuchi} et~al.}{2017}]{2017JPhG...44k4001H}
{Horiuchi} S.,  {Nakamura} K.,  {Takiwaki} T.,   {Kotake} K.,  2017, \mn@doi
  [Journal of Physics G Nuclear Physics] {10.1088/1361-6471/aa8f1f}, 44, 114001

\bibitem[\protect\citeauthoryear{{Horowitz}}{{Horowitz}}{2002}]{2002PhRvD..65d3001H}
{Horowitz} C.~J.,  2002, \mn@doi [\prd] {10.1103/PhysRevD.65.043001}, 65,
  043001

\bibitem[\protect\citeauthoryear{{Horowitz}, {Caballero}, {Lin}, {O'Connor}  \&
  {Schwenk}}{{Horowitz} et~al.}{2017}]{2017PhRvC..95b5801H}
{Horowitz} C.~J.,  {Caballero} O.~L.,  {Lin} Z.,  {O'Connor} E.,   {Schwenk}
  A.,  2017, \mn@doi [\prc] {10.1103/PhysRevC.95.025801}, 95, 025801

\bibitem[\protect\citeauthoryear{{H{\"u}depohl}, {M{\"u}ller}, {Janka}, {Marek}
   \& {Raffelt}}{{H{\"u}depohl} et~al.}{2010}]{2010PhRvL.104y1101H}
{H{\"u}depohl} L.,  {M{\"u}ller} B.,  {Janka} H.~T.,  {Marek} A.,   {Raffelt}
  G.~G.,  2010, \mn@doi [\prl] {10.1103/PhysRevLett.104.251101}, 104, 251101

\bibitem[\protect\citeauthoryear{{Hyper-Kamiokande Proto-Collaboration}
  et~al.,}{{Hyper-Kamiokande Proto-Collaboration}
  et~al.}{2018}]{2018arXiv180504163H}
{Hyper-Kamiokande Proto-Collaboration} et~al., 2018, arXiv e-prints, p.
  arXiv:1805.04163

\bibitem[\protect\citeauthoryear{{Iwakami}, {Okawa}, {Nagakura}, {Harada},
  {Furusawa}, {Sumiyoshi}, {Matsufuru}  \& {Yamada}}{{Iwakami}
  et~al.}{2020}]{2020ApJ...903...82I}
{Iwakami} W.,  {Okawa} H.,  {Nagakura} H.,  {Harada} A.,  {Furusawa} S.,
  {Sumiyoshi} K.,  {Matsufuru} H.,   {Yamada} S.,  2020, \mn@doi [\apj]
  {10.3847/1538-4357/abb8cf}, 903, 82

\bibitem[\protect\citeauthoryear{{Iwakami} et~al.,}{{Iwakami}
  et~al.}{2021}]{2021arXiv210905846I}
{Iwakami} W.,  et~al., 2021, arXiv e-prints, p. arXiv:2109.05846

\bibitem[\protect\citeauthoryear{{Lang}, {McCabe}, {Reichard}, {Selvi}  \&
  {Tamborra}}{{Lang} et~al.}{2016}]{2016PhRvD..94j3009L}
{Lang} R.~F.,  {McCabe} C.,  {Reichard} S.,  {Selvi} M.,   {Tamborra} I.,
  2016, \mn@doi [\prd] {10.1103/PhysRevD.94.103009}, 94, 103009

\bibitem[\protect\citeauthoryear{{Li}, {Roberts}  \& {Beacom}}{{Li}
  et~al.}{2021}]{2021PhRvD.103b3016L}
{Li} S.~W.,  {Roberts} L.~F.,   {Beacom} J.~F.,  2021, \mn@doi [\prd]
  {10.1103/PhysRevD.103.023016}, 103, 023016

\bibitem[\protect\citeauthoryear{{Liebend{\"o}rfer}, {Mezzacappa}, {Messer},
  {Martinez-Pinedo}, {Hix}  \& {Thielemann}}{{Liebend{\"o}rfer}
  et~al.}{2003}]{2003NuPhA.719..144L}
{Liebend{\"o}rfer} M.,  {Mezzacappa} A.,  {Messer} O.~E.~B.,  {Martinez-Pinedo}
  G.,  {Hix} W.~R.,   {Thielemann} F.-K.,  2003, \mn@doi [Nuclear Physics A]
  {10.1016/S0375-9474(03)00984-9}, 719, C144

\bibitem[\protect\citeauthoryear{{Marek}, {Dimmelmeier}, {Janka}, {M{\"u}ller}
  \& {Buras}}{{Marek} et~al.}{2006}]{2006A&A...445..273M}
{Marek} A.,  {Dimmelmeier} H.,  {Janka} H.~T.,  {M{\"u}ller} E.,   {Buras} R.,
  2006, \mn@doi [\aap] {10.1051/0004-6361:20052840}, 445, 273

\bibitem[\protect\citeauthoryear{{Mirizzi}, {Tamborra}, {Janka}, {Saviano},
  {Scholberg}, {Bollig}, {H{\"u}depohl}  \& {Chakraborty}}{{Mirizzi}
  et~al.}{2016}]{2016NCimR..39....1M}
{Mirizzi} A.,  {Tamborra} I.,  {Janka} H.~T.,  {Saviano} N.,  {Scholberg} K.,
  {Bollig} R.,  {H{\"u}depohl} L.,   {Chakraborty} S.,  2016, \mn@doi [Nuovo
  Cimento Rivista Serie] {10.1393/ncr/i2016-10120-8}, 39, 1

\bibitem[\protect\citeauthoryear{{Mori}, {Suwa}, {Nakazato}, {Sumiyoshi},
  {Harada}, {Harada}, {Koshio}  \& {Wendell}}{{Mori}
  et~al.}{2021}]{2021PTEP.2021b3E01M}
{Mori} M.,  {Suwa} Y.,  {Nakazato} K.,  {Sumiyoshi} K.,  {Harada} M.,  {Harada}
  A.,  {Koshio} Y.,   {Wendell} R.~A.,  2021, \mn@doi [Progress of Theoretical
  and Experimental Physics] {10.1093/ptep/ptaa185}, 2021, 023E01

\bibitem[\protect\citeauthoryear{{Morinaga}, {Nagakura}, {Kato}  \&
  {Yamada}}{{Morinaga} et~al.}{2020}]{2020PhRvR...2a2046M}
{Morinaga} T.,  {Nagakura} H.,  {Kato} C.,   {Yamada} S.,  2020, \mn@doi
  [Physical Review Research] {10.1103/PhysRevResearch.2.012046}, 2, 012046

\bibitem[\protect\citeauthoryear{{Morozova}, {Radice}, {Burrows}  \&
  {Vartanyan}}{{Morozova} et~al.}{2018}]{2018ApJ...861...10M}
{Morozova} V.,  {Radice} D.,  {Burrows} A.,   {Vartanyan} D.,  2018, \mn@doi
  [\apj] {10.3847/1538-4357/aac5f1}, 861, 10

\bibitem[\protect\citeauthoryear{{M{\"u}ller} \& {Janka}}{{M{\"u}ller} \&
  {Janka}}{2015}]{2015MNRAS.448.2141M}
{M{\"u}ller} B.,  {Janka} H.-T.,  2015, \mn@doi [\mnras]
  {10.1093/mnras/stv101}, 448, 2141

\bibitem[\protect\citeauthoryear{{M{\"u}ller}, {Melson}, {Heger}  \&
  {Janka}}{{M{\"u}ller} et~al.}{2017}]{2017MNRAS.472..491M}
{M{\"u}ller} B.,  {Melson} T.,  {Heger} A.,   {Janka} H.-T.,  2017, \mn@doi
  [\mnras] {10.1093/mnras/stx1962}, 472, 491

\bibitem[\protect\citeauthoryear{{M{\"u}ller} et~al.,}{{M{\"u}ller}
  et~al.}{2019}]{2019MNRAS.484.3307M}
{M{\"u}ller} B.,  et~al., 2019, \mn@doi [\mnras] {10.1093/mnras/stz216}, 484,
  3307

\bibitem[\protect\citeauthoryear{{Nagakura} et~al.,}{{Nagakura}
  et~al.}{2018}]{2018ApJ...854..136N}
{Nagakura} H.,  et~al., 2018, \mn@doi [\apj] {10.3847/1538-4357/aaac29}, 854,
  136

\bibitem[\protect\citeauthoryear{{Nagakura}, {Furusawa}, {Togashi}, {Richers},
  {Sumiyoshi}  \& {Yamada}}{{Nagakura} et~al.}{2019a}]{2019ApJS..240...38N}
{Nagakura} H.,  {Furusawa} S.,  {Togashi} H.,  {Richers} S.,  {Sumiyoshi} K.,
  {Yamada} S.,  2019a, \mn@doi [\apjs] {10.3847/1538-4365/aafac9}, 240, 38

\bibitem[\protect\citeauthoryear{{Nagakura}, {Takahashi}  \&
  {Yamamoto}}{{Nagakura} et~al.}{2019b}]{2019MNRAS.483..208N}
{Nagakura} H.,  {Takahashi} K.,   {Yamamoto} Y.,  2019b, \mn@doi [\mnras]
  {10.1093/mnras/sty3114}, 483, 208

\bibitem[\protect\citeauthoryear{{Nagakura}, {Sumiyoshi}  \&
  {Yamada}}{{Nagakura} et~al.}{2019c}]{2019ApJ...880L..28N}
{Nagakura} H.,  {Sumiyoshi} K.,   {Yamada} S.,  2019c, \mn@doi [\apjl]
  {10.3847/2041-8213/ab30ca}, 880, L28

\bibitem[\protect\citeauthoryear{{Nagakura}, {Morinaga}, {Kato}  \&
  {Yamada}}{{Nagakura} et~al.}{2019d}]{2019ApJ...886..139N}
{Nagakura} H.,  {Morinaga} T.,  {Kato} C.,   {Yamada} S.,  2019d, \mn@doi
  [\apj] {10.3847/1538-4357/ab4cf2}, 886, 139

\bibitem[\protect\citeauthoryear{{Nagakura}, {Burrows}, {Radice}  \&
  {Vartanyan}}{{Nagakura} et~al.}{2020}]{2020MNRAS.492.5764N}
{Nagakura} H.,  {Burrows} A.,  {Radice} D.,   {Vartanyan} D.,  2020, \mn@doi
  [\mnras] {10.1093/mnras/staa261}, 492, 5764

\bibitem[\protect\citeauthoryear{{Nagakura}, {Burrows}, {Johns}  \&
  {Fuller}}{{Nagakura} et~al.}{2021a}]{2021PhRvD.104h3025N}
{Nagakura} H.,  {Burrows} A.,  {Johns} L.,   {Fuller} G.~M.,  2021a, \mn@doi
  [\prd] {10.1103/PhysRevD.104.083025}, 104, 083025

\bibitem[\protect\citeauthoryear{{Nagakura}, {Burrows}, {Vartanyan}  \&
  {Radice}}{{Nagakura} et~al.}{2021b}]{2021MNRAS.500..696N}
{Nagakura} H.,  {Burrows} A.,  {Vartanyan} D.,   {Radice} D.,  2021b, \mn@doi
  [\mnras] {10.1093/mnras/staa2691}, 500, 696

\bibitem[\protect\citeauthoryear{{Nagakura}, {Burrows}  \&
  {Vartanyan}}{{Nagakura} et~al.}{2021c}]{2021MNRAS.506.1462N}
{Nagakura} H.,  {Burrows} A.,   {Vartanyan} D.,  2021c, \mn@doi [\mnras]
  {10.1093/mnras/stab1785}, 506, 1462

\bibitem[\protect\citeauthoryear{{Nakazato} \& {Suzuki}}{{Nakazato} \&
  {Suzuki}}{2019}]{2019ApJ...878...25N}
{Nakazato} K.,  {Suzuki} H.,  2019, \mn@doi [\apj] {10.3847/1538-4357/ab1d4b},
  878, 25

\bibitem[\protect\citeauthoryear{{Nakazato}, {Sumiyoshi}, {Suzuki}, {Totani},
  {Umeda}  \& {Yamada}}{{Nakazato} et~al.}{2013}]{2013ApJS..205....2N}
{Nakazato} K.,  {Sumiyoshi} K.,  {Suzuki} H.,  {Totani} T.,  {Umeda} H.,
  {Yamada} S.,  2013, \mn@doi [\apjs] {10.1088/0067-0049/205/1/2}, 205, 2

\bibitem[\protect\citeauthoryear{{O'Connor} \& {Couch}}{{O'Connor} \&
  {Couch}}{2018}]{2018ApJ...865...81O}
{O'Connor} E.~P.,  {Couch} S.~M.,  2018, \mn@doi [\apj]
  {10.3847/1538-4357/aadcf7}, 865, 81

\bibitem[\protect\citeauthoryear{{O'Connor} \& {Ott}}{{O'Connor} \&
  {Ott}}{2013}]{2013ApJ...762..126O}
{O'Connor} E.,  {Ott} C.~D.,  2013, \mn@doi [\apj]
  {10.1088/0004-637X/762/2/126}, 762, 126

\bibitem[\protect\citeauthoryear{{Obergaulinger} \& {Aloy}}{{Obergaulinger} \&
  {Aloy}}{2021}]{2021arXiv210813864O}
{Obergaulinger} M.,  {Aloy} M.~{\'A}.,  2021, arXiv e-prints, p.
  arXiv:2108.13864

\bibitem[\protect\citeauthoryear{{Obergaulinger}, {Aloy}  \&
  {M{\"u}ller}}{{Obergaulinger} et~al.}{2006}]{2006A&A...450.1107O}
{Obergaulinger} M.,  {Aloy} M.~A.,   {M{\"u}ller} E.,  2006, \mn@doi [\aap]
  {10.1051/0004-6361:20054306}, 450, 1107

\bibitem[\protect\citeauthoryear{{Ott}, {Burrows}, {Livne}  \& {Walder}}{{Ott}
  et~al.}{2004}]{2004ApJ...600..834O}
{Ott} C.~D.,  {Burrows} A.,  {Livne} E.,   {Walder} R.,  2004, \mn@doi [\apj]
  {10.1086/379822}, 600, 834

\bibitem[\protect\citeauthoryear{{Pajkos}, {Couch}, {Pan}  \&
  {O'Connor}}{{Pajkos} et~al.}{2019}]{2019ApJ...878...13P}
{Pajkos} M.~A.,  {Couch} S.~M.,  {Pan} K.-C.,   {O'Connor} E.~P.,  2019,
  \mn@doi [\apj] {10.3847/1538-4357/ab1de2}, 878, 13

\bibitem[\protect\citeauthoryear{{Pattavina}, {Ferreiro Iachellini}  \&
  {Tamborra}}{{Pattavina} et~al.}{2020}]{2020PhRvD.102f3001P}
{Pattavina} L.,  {Ferreiro Iachellini} N.,   {Tamborra} I.,  2020, \mn@doi
  [\prd] {10.1103/PhysRevD.102.063001}, 102, 063001

\bibitem[\protect\citeauthoryear{{Pattavina} et~al.,}{{Pattavina}
  et~al.}{2021}]{2021JCAP...10..064P}
{Pattavina} L.,  et~al., 2021, \mn@doi [\jcap] {10.1088/1475-7516/2021/10/064},
  2021, 064

\bibitem[\protect\citeauthoryear{{Pons}, {Reddy}, {Prakash}, {Lattimer}  \&
  {Miralles}}{{Pons} et~al.}{1999}]{1999ApJ...513..780P}
{Pons} J.~A.,  {Reddy} S.,  {Prakash} M.,  {Lattimer} J.~M.,   {Miralles}
  J.~A.,  1999, \mn@doi [\apj] {10.1086/306889}, 513, 780

\bibitem[\protect\citeauthoryear{{Powell} \& {M{\"u}ller}}{{Powell} \&
  {M{\"u}ller}}{2019}]{2019MNRAS.487.1178P}
{Powell} J.,  {M{\"u}ller} B.,  2019, \mn@doi [\mnras] {10.1093/mnras/stz1304},
  487, 1178

\bibitem[\protect\citeauthoryear{{Reed}, {Fattoyev}, {Horowitz}  \&
  {Piekarewicz}}{{Reed} et~al.}{2021}]{2021PhRvL.126q2503R}
{Reed} B.~T.,  {Fattoyev} F.~J.,  {Horowitz} C.~J.,   {Piekarewicz} J.,  2021,
  \mn@doi [\prl] {10.1103/PhysRevLett.126.172503}, 126, 172503

\bibitem[\protect\citeauthoryear{{Richers}, {Ott}, {Abdikamalov}, {O'Connor}
  \& {Sullivan}}{{Richers} et~al.}{2017}]{2017PhRvD..95f3019R}
{Richers} S.,  {Ott} C.~D.,  {Abdikamalov} E.,  {O'Connor} E.,   {Sullivan} C.,
   2017, \mn@doi [\prd] {10.1103/PhysRevD.95.063019}, 95, 063019

\bibitem[\protect\citeauthoryear{{Roberts}, {Shen}, {Cirigliano}, {Pons},
  {Reddy}  \& {Woosley}}{{Roberts} et~al.}{2012}]{2012PhRvL.108f1103R}
{Roberts} L.~F.,  {Shen} G.,  {Cirigliano} V.,  {Pons} J.~A.,  {Reddy} S.,
  {Woosley} S.~E.,  2012, \mn@doi [\prl] {10.1103/PhysRevLett.108.061103}, 108,
  061103

\bibitem[\protect\citeauthoryear{{Scheidegger}, {Fischer}, {Whitehouse}  \&
  {Liebend{\"o}rfer}}{{Scheidegger} et~al.}{2008}]{2008A&A...490..231S}
{Scheidegger} S.,  {Fischer} T.,  {Whitehouse} S.~C.,   {Liebend{\"o}rfer} M.,
  2008, \mn@doi [\aap] {10.1051/0004-6361:20078577}, 490, 231

\bibitem[\protect\citeauthoryear{{Scholberg}}{{Scholberg}}{2012}]{2012ARNPS..62...81S}
{Scholberg} K.,  2012, \mn@doi [Annual Review of Nuclear and Particle Science]
  {10.1146/annurev-nucl-102711-095006}, 62, 81

\bibitem[\protect\citeauthoryear{{Sekiguchi}}{{Sekiguchi}}{2010}]{2010PThPh.124..331S}
{Sekiguchi} Y.,  2010, \mn@doi [Progress of Theoretical Physics]
  {10.1143/PTP.124.331}, 124, 331

\bibitem[\protect\citeauthoryear{{Shalgar} \& {Tamborra}}{{Shalgar} \&
  {Tamborra}}{2019}]{2019ApJ...883...80S}
{Shalgar} S.,  {Tamborra} I.,  2019, \mn@doi [\apj] {10.3847/1538-4357/ab38ba},
  883, 80

\bibitem[\protect\citeauthoryear{{Shibata} \& {Sekiguchi}}{{Shibata} \&
  {Sekiguchi}}{2004}]{2004PhRvD..69h4024S}
{Shibata} M.,  {Sekiguchi} Y.-I.,  2004, \mn@doi [\prd]
  {10.1103/PhysRevD.69.084024}, \href
  {https://ui.adsabs.harvard.edu/abs/2004PhRvD..69h4024S} {69, 084024}

\bibitem[\protect\citeauthoryear{{Skinner}, {Dolence}, {Burrows}, {Radice}  \&
  {Vartanyan}}{{Skinner} et~al.}{2019}]{2019ApJS..241....7S}
{Skinner} M.~A.,  {Dolence} J.~C.,  {Burrows} A.,  {Radice} D.,   {Vartanyan}
  D.,  2019, \mn@doi [The Astrophysical Journal Supplement Series]
  {10.3847/1538-4365/ab007f}, 241, 7

\bibitem[\protect\citeauthoryear{{Sotani}, {Kuroda}, {Takiwaki}  \&
  {Kotake}}{{Sotani} et~al.}{2017}]{2017PhRvD..96f3005S}
{Sotani} H.,  {Kuroda} T.,  {Takiwaki} T.,   {Kotake} K.,  2017, \mn@doi [\prd]
  {10.1103/PhysRevD.96.063005}, 96, 063005

\bibitem[\protect\citeauthoryear{{Sotani}, {Takiwaki}  \& {Togashi}}{{Sotani}
  et~al.}{2021}]{2021arXiv211003131S}
{Sotani} H.,  {Takiwaki} T.,   {Togashi} H.,  2021, arXiv e-prints, p.
  arXiv:2110.03131

\bibitem[\protect\citeauthoryear{{Steiner}, {Hempel}  \& {Fischer}}{{Steiner}
  et~al.}{2013}]{2013ApJ...774...17S}
{Steiner} A.~W.,  {Hempel} M.,   {Fischer} T.,  2013, \mn@doi [\apj]
  {10.1088/0004-637X/774/1/17}, 774, 17

\bibitem[\protect\citeauthoryear{{Sukhbold}, {Woosley}  \& {Heger}}{{Sukhbold}
  et~al.}{2018}]{2018ApJ...860...93S}
{Sukhbold} T.,  {Woosley} S.~E.,   {Heger} A.,  2018, \mn@doi [\apj]
  {10.3847/1538-4357/aac2da}, 860, 93

\bibitem[\protect\citeauthoryear{{Sumiyoshi}, {Nakazato}, {Suzuki}, {Hu}  \&
  {Shen}}{{Sumiyoshi} et~al.}{2019}]{2019ApJ...887..110S}
{Sumiyoshi} K.,  {Nakazato} K.,  {Suzuki} H.,  {Hu} J.,   {Shen} H.,  2019,
  \mn@doi [\apj] {10.3847/1538-4357/ab5443}, 887, 110

\bibitem[\protect\citeauthoryear{{Summa}, {Janka}, {Melson}  \&
  {Marek}}{{Summa} et~al.}{2018}]{2018ApJ...852...28S}
{Summa} A.,  {Janka} H.-T.,  {Melson} T.,   {Marek} A.,  2018, \mn@doi [\apj]
  {10.3847/1538-4357/aa9ce8}, 852, 28

\bibitem[\protect\citeauthoryear{{Suwa}, {Yoshida}, {Shibata}, {Umeda}  \&
  {Takahashi}}{{Suwa} et~al.}{2015}]{2015MNRAS.454.3073S}
{Suwa} Y.,  {Yoshida} T.,  {Shibata} M.,  {Umeda} H.,   {Takahashi} K.,  2015,
  \mn@doi [\mnras] {10.1093/mnras/stv2195}, 454, 3073

\bibitem[\protect\citeauthoryear{{Suwa}, {Harada}, {Nakazato}  \&
  {Sumiyoshi}}{{Suwa} et~al.}{2021}]{2021PTEP.2021a3E01S}
{Suwa} Y.,  {Harada} A.,  {Nakazato} K.,   {Sumiyoshi} K.,  2021, \mn@doi
  [Progress of Theoretical and Experimental Physics] {10.1093/ptep/ptaa154},
  2021, 013E01

\bibitem[\protect\citeauthoryear{{Takiwaki} \& {Kotake}}{{Takiwaki} \&
  {Kotake}}{2011}]{2011ApJ...743...30T}
{Takiwaki} T.,  {Kotake} K.,  2011, \mn@doi [\apj]
  {10.1088/0004-637X/743/1/30}, 743, 30

\bibitem[\protect\citeauthoryear{{Tamborra}, {Hanke}, {Janka}, {M{\"u}ller},
  {Raffelt}  \& {Marek}}{{Tamborra} et~al.}{2014}]{2014ApJ...792...96T}
{Tamborra} I.,  {Hanke} F.,  {Janka} H.-T.,  {M{\"u}ller} B.,  {Raffelt} G.~G.,
    {Marek} A.,  2014, \mn@doi [\apj] {10.1088/0004-637X/792/2/96}, 792, 96

\bibitem[\protect\citeauthoryear{{Tews}, {Lattimer}, {Ohnishi}  \&
  {Kolomeitsev}}{{Tews} et~al.}{2017}]{2017ApJ...848..105T}
{Tews} I.,  {Lattimer} J.~M.,  {Ohnishi} A.,   {Kolomeitsev} E.~E.,  2017,
  \mn@doi [\apj] {10.3847/1538-4357/aa8db9}, 848, 105

\bibitem[\protect\citeauthoryear{{Thompson}, {Burrows}  \& {Pinto}}{{Thompson}
  et~al.}{2003}]{2003ApJ...592..434T}
{Thompson} T.~A.,  {Burrows} A.,   {Pinto} P.~A.,  2003, \mn@doi [\apj]
  {10.1086/375701}, 592, 434

\bibitem[\protect\citeauthoryear{{Torres-Forn{\'e}}, {Cerd{\'a}-Dur{\'a}n},
  {Obergaulinger}, {M{\"u}ller}  \& {Font}}{{Torres-Forn{\'e}}
  et~al.}{2019}]{2019PhRvL.123e1102T}
{Torres-Forn{\'e}} A.,  {Cerd{\'a}-Dur{\'a}n} P.,  {Obergaulinger} M.,
  {M{\"u}ller} B.,   {Font} J.~A.,  2019, \mn@doi [\prl]
  {10.1103/PhysRevLett.123.051102}, 123, 051102

\bibitem[\protect\citeauthoryear{{Vartanyan}, {Burrows}, {Radice}, {Skinner}
  \& {Dolence}}{{Vartanyan} et~al.}{2019a}]{2019MNRAS.482..351V}
{Vartanyan} D.,  {Burrows} A.,  {Radice} D.,  {Skinner} M.~A.,   {Dolence} J.,
  2019a, \mn@doi [\mnras] {10.1093/mnras/sty2585}, 482, 351

\bibitem[\protect\citeauthoryear{{Vartanyan}, {Burrows}  \&
  {Radice}}{{Vartanyan} et~al.}{2019b}]{2019MNRAS.489.2227V}
{Vartanyan} D.,  {Burrows} A.,   {Radice} D.,  2019b, \mn@doi [\mnras]
  {10.1093/mnras/stz2307}, 489, 2227

\bibitem[\protect\citeauthoryear{{Vartanyan}, {Coleman}  \&
  {Burrows}}{{Vartanyan} et~al.}{2021a}]{2021arXiv210910920V}
{Vartanyan} D.,  {Coleman} M. S.~B.,   {Burrows} A.,  2021a, arXiv e-prints, p.
  arXiv:2109.10920

\bibitem[\protect\citeauthoryear{{Vartanyan}, {Laplace}, {Renzo},
  {G{\"o}tberg}, {Burrows}  \& {de Mink}}{{Vartanyan}
  et~al.}{2021b}]{2021ApJ...916L...5V}
{Vartanyan} D.,  {Laplace} E.,  {Renzo} M.,  {G{\"o}tberg} Y.,  {Burrows} A.,
  {de Mink} S.~E.,  2021b, \mn@doi [\apjl] {10.3847/2041-8213/ac0b42}, 916, L5

\bibitem[\protect\citeauthoryear{{Wallace}, {Burrows}  \& {Dolence}}{{Wallace}
  et~al.}{2016}]{2016ApJ...817..182W}
{Wallace} J.,  {Burrows} A.,   {Dolence} J.~C.,  2016, \mn@doi [\apj]
  {10.3847/0004-637X/817/2/182}, 817, 182

\bibitem[\protect\citeauthoryear{{Warren}, {Couch}, {O'Connor}  \&
  {Morozova}}{{Warren} et~al.}{2020}]{2020ApJ...898..139W}
{Warren} M.~L.,  {Couch} S.~M.,  {O'Connor} E.~P.,   {Morozova} V.,  2020,
  \mn@doi [\apj] {10.3847/1538-4357/ab97b7}, 898, 139

\bibitem[\protect\citeauthoryear{{Wei}, {Burgio}  \& {Schulze}}{{Wei}
  et~al.}{2019}]{2019MNRAS.484.5162W}
{Wei} J.~B.,  {Burgio} G.~F.,   {Schulze} H.~J.,  2019, \mn@doi [\mnras]
  {10.1093/mnras/stz336}, 484, 5162

\bibitem[\protect\citeauthoryear{{Wongwathanarat}, {Janka}  \&
  {M{\"u}ller}}{{Wongwathanarat} et~al.}{2010}]{2010ApJ...725L.106W}
{Wongwathanarat} A.,  {Janka} H.-T.,   {M{\"u}ller} E.,  2010, \mn@doi [\apjl]
  {10.1088/2041-8205/725/1/L106}, 725, L106

\bibitem[\protect\citeauthoryear{{Yakovlev}, {Kaminker}, {Gnedin}  \&
  {Haensel}}{{Yakovlev} et~al.}{2001}]{2001PhR...354....1Y}
{Yakovlev} D.~G.,  {Kaminker} A.~D.,  {Gnedin} O.~Y.,   {Haensel} P.,  2001,
  \mn@doi [\physrep] {10.1016/S0370-1573(00)00131-9}, 354, 1

\bibitem[\protect\citeauthoryear{{Yoshida}, {Takiwaki}, {Kotake}, {Takahashi},
  {Nakamura}  \& {Umeda}}{{Yoshida} et~al.}{2021}]{2021ApJ...908...44Y}
{Yoshida} T.,  {Takiwaki} T.,  {Kotake} K.,  {Takahashi} K.,  {Nakamura} K.,
  {Umeda} H.,  2021, \mn@doi [\apj] {10.3847/1538-4357/abd3a3}, 908, 44

\bibitem[\protect\citeauthoryear{{Young} et~al.,}{{Young}
  et~al.}{2006}]{2006ApJ...640..891Y}
{Young} P.~A.,  et~al., 2006, \mn@doi [\apj] {10.1086/500108}, 640, 891

\makeatother
\end{thebibliography}







\bsp	
\label{lastpage}
\end{document}